# ATOMIC TRANSITION PROBABILITIES FOR UV AND OPTICAL LINES OF Tm II[1]

(Short Title: TRANSITION PROBABILITIES OF Tm II)

E. A. Den Hartog[1], G. T. Voith[1], and I. U. Roederer[2]

[1]Department of Physics, University of Wisconsin – Madison, 1150 University Ave, Madison, WI 53706; eadenhar@wisc.edu; gtvoith@wisc.edu
[2]Department of Physics, North Carolina State University, Raleigh, NC 27695; iuroederer@ncsu.edu

ORCIDS:

E. A. Den Hartog:   0000-0001-8582-0910

I. U. Roederer      0000-0001-5107-8930

ABSTRACT

We report new branching fraction measurements for 224 ultraviolet (UV) and optical transitions of Tm II. These transitions range in wavelength (wavenumber) from 2350 – 6417 Å (42532 – 15579 cm$^{-1}$) and originate in 13 odd-parity and 24 even-parity upper levels. Thirty-five of the 37 levels, accounting for 213 of the 224 transitions, are studied for the first time. Branching fractions are determined for two levels studied previously for comparison to earlier results. The levels studied for the first time are high-lying, ranging in energy from 35753 – 54989 cm$^{-1}$. The branching fractions are determined from emission spectra from two different high-resolution spectrometers. These are combined with radiative lifetimes reported in an earlier study to produce a set of transition probabilities and log($gf$) values with accuracy ranging from 5 – 30%. Comparison is made to experimental and theoretical transition probabilities from the literature where such data exist. These new log($gf$) values are used to derive an abundance from one previously unused Tm II line in the UV spectrum of the r-process-enhanced metal-poor star HD 222925, and this abundance is consistent with previous determinations based on other Tm II lines.

1. INTRODUCTION

Elements all across the periodic table were formed by the earliest generations of stars. Open questions remain about the physical mechanisms of the nucleosynthesis processes that produce these elements and the sites where they occur. For example, the Solar abundances of the elements listed along the bottom two-thirds of the periodic table (atomic number $Z > 30$) can be mostly accounted for by two general mechanisms, the rapid (r) and slow (s) neutron-capture processes (Sneden et al. 2008). S-process nucleosynthesis is generally understood to have occurred in low- and intermediate-mass stars during thermal pulses that occur while these stars undergo double shell burning near the end of their lives (e.g., Karakas 2010), and in massive stars that rotate rapidly (e.g., Frischknecht et al. 2016). In contrast, r-process nucleosynthesis has only been definitively associated with one astrophysical site, a merging pair of neutron stars (e.g., Drout et al. 2017; Cowan et al. 2021).

The lanthanide elements ($57 \leq Z \leq 71$) have played an outsized role in confirming these associations and characterizing the properties of the sites. Some of the lanthanides, such as cerium and neodymium, are relatively easy to detect in spectra of the atmospheres of cool stars because their dominant species, the first ionization state, present tens or hundreds of absorption lines in the optical and ultraviolet (UV) spectral range ($\approx$ 2000 – 10000 Å; e.g., Moore et al. 1966). Others, such as europium and ytterbium, present few but strong optical transitions that are relatively easy to detect (e.g., Sneden et al. 2009), even when the elemental abundance is low (e.g., Honda et al. 2006). Lanthanide abundances in old, metal-poor stars also indicate the relative contributions of material produced by the r- or s-processes in prior generations of stars (e.g., Simmerer et al. 2004), the physics responsible for their production (e.g., Roederer et al. 2023), the chemical evolution that occurred in our Galaxy (e.g., Magrini et al. 2018), the galactic components that assembled our Galaxy (e.g., Gull et al. 2021), and the heat budgets of exoplanets (e.g., Wang et al. 2020). Lanthanides also dominate the late-time opacity of the ejecta from merging neutron stars (e.g., Kasen et al. 2013; Tanvir et al. 2017), providing the key evidence that directly links r-process nucleosynthesis to this site.

All of these applications require accurate transition probabilities for lanthanide elements across a wide range of wavelengths and level energies. In most modern laboratory studies, transition probabilities (Einstein $A$-values) are determined by combining branching fractions (BFs) with the radiative lifetime of each upper level. The BFs are normalized relative intensities of transitions connected to a given upper level and are usually determined from high-resolution emission spectra. The radiative lifetime of the upper level is often determined using time-resolved laser-induced fluorescence (TRLIF) and is used to put the BFs on an absolute scale. The current study follows this methodology.

Thulium (Th, $Z= 69$) is a lanthanide element which has been the focus of some laboratory work, but there is still more to do. The literature on modern experimental studies of transition probabilities for Tm II consist of several papers on radiative lifetimes and two publications on transition probabilities. The most comprehensive study was carried out in our University of Wisconsin — Madison (UW) group. Anderson, Den Hartog & Lawler (1996, hereafter UW96) used TRLIF to measure radiative lifetimes for 85 even-parity and 19 odd-parity levels of Tm II ranging in energy from 23000 to 55000 cm$^{-1}$, as well as radiative lifetimes for a total of 194 levels of Tm I. A study of transition probabilities followed. Wickliffe & Lawler (1997, hereafter UW97) used high resolution emission spectra of hollow cathode lamps (HCLs) recorded with a 1-m Fourier Transform Spectrometer (FTS) to determine BFs for transitions associated with levels having radiative lifetimes from UW96. (See also Table 11 in the appendix of Lawler et al. 2009 for a machine-readable table of the UW97 results) These BFs and lifetimes were combined to determine *A*-values and log(*gf*)s for 146 transitions from 56 levels of Tm II ranging in energy from 23800 – 47300 cm$^{-1}$. In addition, transition probabilities were determined for 376 transitions of Tm I. Radiative lifetimes were also reported by Rieger, McCurdy & Pinnington (1999), who used the more precise fast-beam-laser technique to re-measure 11 of the lifetimes of UW96 to resolve differences between that work and the relativistic theory of Quinet, Palmeri & Biémont (1999). Their work supported the lifetimes of UW96 except for one level which was discrepant by ~2.5 times the combined uncertainties. Xu, Jiang & Svanberg (2003) reported TRLIF lifetimes for that level and one other in Tm II, as well as 8 in Tm I and 3 in Tm III. Their Tm II lifetimes agreed with UW96 lifetimes within their uncertainties and not with Rieger et al. for the discrepant level. Finally, Tian et al. (2016) measured lifetimes using TRLIF for 88 levels of Tm I and 29 levels of Tm II. Their Tm II lifetimes overlapped with the UW96 study for 7 levels, for which good agreement was observed. Recently Wang et al. (2022) published a study of BFs for 80 transitions of Tm I and 30 transitions of Tm II, combining them with the lifetimes of Tian et al. and one lifetime from UW96 to determine transition probabilities.

In addition to these experimental studies, Quinet, Palmeri & Biémont (1999) used the semi-empirical Relativistic Hartree-Fock method that included configuration interaction and core-polarization effects to calculate transition probabilities for 352 transitions of Tm II. A more extended list is available for download from the D.R.E.A.M. database (Quinet & Palmeri 2020), and includes 7881 transitions. Quinet et al. made detailed comparison to the radiative lifetimes of UW96 and the transition probabilities of UW97. There is also a more recent theoretical study that used the ab initio multiconfiguration Dirac-Hartree-Fock and relativistic configuration interaction methods as implemented in the GRASP2018 package. Radžiūtė et al. (2021) reported calculated energy levels, wavelengths and *A*-values for millions of transitions for the singly-ionized rare earths Z=65–70. This study was aimed at providing the massive amount of transition data required to calculate the lanthanide opacity following merging neutron stars.

Of the 104 Tm II levels that have radiative lifetime measurements in the UW96 study, UW97 reported *A*-values and log(*gf*)s for transitions associated with 56 of them. Some of the remaining 48 levels were not included because of troublesome blends in one or more transition, but many high-lying levels were eliminated from the study because they had one or more significant transition below the ~2900 Å limit of their FTS spectra. In 2012, a powerful new high-resolution spectrograph was added to the UW laboratory. The UW 3-m High-Resolution Echelle Spectrograph has excellent UV capability, and has opened up the possibility of studying high-lying levels having significant UV branches. Of the high-lying levels of Tm II omitted by UW97, we have determined BFs for 35 new levels in the current study.

In section 2 below, we describe our BF measurements, including the two spectrometers used and the relative radiometric calibration of each, methods for spectral analysis and for dealing with blends. We present our results and make comparison to experiment and theory from the literature in section 3. Finally, in section 4 we employ the new data to identify line(s) that may be of use as Tm II abundance indicators in metal-poor stars.

## 2. BRANCHING FRACTIONS OF Tm II

The BF for a transition between an upper level $u$ and lower level $l$ is the ratio of its $A$-value to the sum of the $A$-values associated with $u$. This can also be expressed as the ratio of relative emission intensities $I$ (in any units proportional to photons/time) for these transitions:

$$BF_{ul} = \frac{A_{ul}}{\sum_l A_{ul}} = \frac{I_{ul}}{\sum_l I_{ul}}, \qquad (1)$$

where the sum in the denominator is over all transitions associated with level $u$. The BFs associated with $u$, by definition, sum to one. It is therefore important when measuring BFs to account for all possible decay paths from an upper level so that the normalization is correct. (Relative intensities where the sum is over less than the full complement of transitions are referred to as branching ratios (BRs).) The energy level structure of $Tm^+$ is sufficiently well known[2] for this task and there are few missing low-lying levels of either parity that might give rise to missing branches in the current study. All possible dipole-allowed transitions obeying the ΔJ and parity change selection rules are investigated and all observed transitions are analyzed. Some weak transitions that have >30% final uncertainty are not included in the tables, but are included in the normalization. The sum of these "residual" transitions ranges between 0 – 10% for the levels in this study.

### 2.1 Branching Ratios from Two High-Resolution Spectrometers

To cover the full wavelength range of transitions from these high-lying levels we have employed two high resolution spectrometers. Spectra from the archives of the 1 m Fourier Transform

---

[2] Throughout this manuscript and associated tables, we use the energy levels from Martin, Zalubus & Hagan (1978) downloaded from the National Institute of Science and Technology Atomic Spectra Database (NIST ASD) (Kramida et al. 2022, available at https://www.nist.gov/pml/atomic-spectra-database) with additional levels and some corrected J values from the work of Wyart (2011). Transition wavenumbers are calculated from the difference in level energies and these are converted to air wavelengths using the standard index of air (Peck & Reeder 1972).

Spectrometer[3] (FTS) on the McMath Telescope at the National Solar Observatory, Kitt Peak, AZ were used to measure BRs for transitions spanning from ~2900 Å in the near-UV through near-infrared wavelengths and the UW 3-m Echelle Spectrograph for the transitions in the range 2350 – 4360 Å. BRs are determined for the transitions covered by each instrument, with one to several near-UV lines acting as a bridge to combine the two sets of BRs. This combined set is then renormalized to determine BFs.

FTSs have many characteristics that make them advantageous for measuring BFs, including excellent absolute wavenumber accuracy, broad spectral coverage and high resolving power. A broad spectrum covering the UV to near-infrared can be recorded in just a few minutes. As an interferometric device, all spectral elements are recorded simultaneously with the result that small drifts in source current do not result in relative intensity errors. The main drawback of the instrument, multiplex noise, also results from interferometry. Multiplex noise is the Poisson statistical noise arising from all lines in the spectrum and is spread evenly throughout the spectrum. As a result, the noise arising from very strong lines in the spectrum will tend to overwhelm the weaker lines. If the source current is increased to bring the weaker branches out of the noise, care must be taken to determine whether the strong branches to low-lying levels remain optically thin.

The UW 3-m Echelle Spectrograph is described in detail in Wood & Lawler (2012). The instrument has a large (128 × 254 mm ruled area,) coarse (23.2 grooves mm$^{-1}$) echelle grating blazed at 63.5°. It operates in very high orders (up to order $m$=385 at 2000 Å), resulting in a high resolving power (up to ~300,000) and has good sensitivity down to 2000 Å. The typical mode of operation for branching fraction work is to utilize a 50 $\mu$m entrance pinhole. The exit plane is imaged onto a 2048 × 2048 CCD detector after passing through a prismatic order separator. The advantage of this instrument is that, as a dispersive instrument, it is free of multiplex noise and can be used to measure very weak branches with good S/N even while keeping the source current at a modest level to avoid self-absorption on strong branches. It has good resolving power but cannot match the resolving power or absolute wavenumber accuracy of the FTS. One disadvantage of this spectrometer arises from its high dispersion. Like the FTS, all spectral elements are collected simultaneously on a given CCD frame, however it requires multiple CCD frames overlapping in the high-dispersion direction to capture the entire blaze envelope from each grating order. In the UV three frames are required, but we typically collect four or five for some redundancy. These must be combined into one spectrum by comparing the intensities of lines that appear in the region where pairs of frames overlap.

The FTS spectra used in this study are listed in Table 1 and are a subset of the spectra used in the earlier UW97 study of Tm II. The 3-m echelle spectra are listed in Table 2. Tm-Ar and Tm-Ne commercial sealed HCLs are the line sources for all spectra except for the FTS spectra with indices 10 and 11 in Table 1. These two spectra were taken with a water-cooled demountable HCL operating at high currents. The Tm-Ne FTS spectra listed at the end of Table 1 have much inferior S/N compared to the Tm-Ar spectra and were used only in the evaluation of potential Ar blends. The commercial HCLs were operated with forced air cooling well above the maximum rated current from the manufacturer in order to bring up weaker lines in the spectra. A range of currents was used to check for evidence of self-absorption on the strongest lines. No self-absorption was observed in the transitions involved in this study. One unintended consequence of the higher lamp currents is that in addition to Tm and buffer gas lines, lines of Fe I and Fe II are also observed in the spectra of the commercial lamps. This is a result of the cathode construction, which is not pure Tm, but rather a thin lining of Tm inside a cylindrical Fe shell. For the most part, the presence of Fe lines is of little consequence to our analysis, but they do contribute to the multiplex noise as well as resulting in additional blend possibilities that must be considered.

---

[3] This instrument was decommissioned in 2012, but all spectra recorded with it are archived and publicly available at https://nispdata.nso.edu/ftp/FTS_cdrom/.

Analysis of the FTS data is done with interactive software written in-house and is much the same as that used in the earlier UW97 study, although it has evolved over the intervening 25 years. The software automatically looks for all possible $u \rightarrow l$ transitions that satisfy the parity change and $\Delta J$ selection rules, displaying a portion of the spectrum centered on each. The positions for the line of interest as well as other possible Tm I, Tm II, buffer gas and contaminant lines are indicated on the plot. If the desired line is observed, the user then interactively sets the baseline and integration limits and a numerical integration is performed to determine the raw intensity. These cannot be turned into BRs or BFs, however, without first determining the relative radiometric calibration for the spectrum. The FTS spectra are calibrated by measuring ratios of line intensities for sets of Ar I & II lines and comparing them to well-known BRs that have been measured for this purpose independently by Whaling, Carle & Pitt (1993), Hashiguchi & Hasikuni (1985) and Danzmann & Kock (1982). Using these internal standards has the advantage that the calibration line light follows the exact same trajectory and encounters the same windows and optics as the Tm II line light, so that effects such as the variation of window transmittance or reflection off the back of the cathode are automatically accounted for in the calibration. The same software and methods are used for integrating the Ar I & II lines as for the Tm II lines. The raw intensity of these calibration lines divided by their known BRs results in overlapping subsets of points that can be used to construct a relative sensitivity versus wavenumber curve.

The 3-m echelle data is analyzed in much the same way except the spectrum is displayed as a 2D perspective plot. The user can interactively cut away rows of pixels running along the high-dispersion direction if needed to isolate the line of interest from lines in adjacent orders. The remaining pixels are summed in the low-dispersion direction to make a 1D high-dispersion spectrum. The analysis then proceeds the same as the FTS spectral analysis with the setting of background and integration limits followed by numerical integration across the line. After the line is integrated, a $D_2$ lamp calibration spectrum, which was recorded immediately following the

HCL spectrum, is displayed and an integral is performed across the grating order at the same spectral location as the line integral. This integral divided by the lamp irradiance yields the relative sensitivity of the instrument and captures both the slow variation of sensitivity in the low-dispersion direction as well as the rapid variation of the grating blaze envelope in the high-dispersion direction.

A typical way to use a continuum lamp for calibrating an instrument would be to have two continuum lamps – one everyday lamp for which the calibrated irradiance inevitably degrades over time due to lamp aging and UV damage to the window, and one little-used lamp that is assumed to have a stable irradiance calibration that is periodically transferred onto the everyday lamp. $D_2$ lamps are typically only calibrated to 4000 Å and that calibration is made at low resolution. The measured calibration irradiance will therefore include irradiance from both the continuum and an increasingly dense forest of lines between 3700 and 4000 Å. To use such a lamp calibration at high resolution requires estimating and applying corrections to the calibration above 3700 Å to remove the effects of the line radiation. We have recently moved away from this practice and instituted a detector-based calibration instead. The relative irradiance of the everyday $D_2$ lamp has been measured in our lab using a NIST-calibrated photodiode detector and a Hg pen lamp source. This method is described in detail in Den Hartog et al. (2023) and the reader is referred to that paper for methodology. The advantage of a detector-based calibration is that, unlike the $D_2$ lamp calibration, the detector calibration should remain stable for many years. The Hg pen lamp + NIST photodiode combination also allows a calibration out to 4360 Å. The sensitivity of the 3-m echelle is determined by dividing the measured $D_2$ intensity by the calibrated relative irradiance of the lamp.

The systematic uncertainty of the instrument calibration is conservatively estimated to be 0.001% per $cm^{-1}$ between the line of interest and the dominant line(s) from the upper level. When there is a single dominant line, the cumulative uncertainty of the calibration is attributed entirely to the weak branches. In the case where there are multiple strong branches, they share this calibration uncertainty between them in proportion to the inverse of the BFs. For example, two lines, both with 0.5 BF and separated by 8000 $cm^{-1}$, would each have 4% calibration uncertainty ascribed to them. By contrast, two lines with 0.25 and 0.75 BFs separated by 8000 $cm^{-1}$ would have 6% and 2% calibration uncertainty, respectively.

The raw intensity of Tm II lines can be converted to BRs by dividing by the relative sensitivity of the instrument. After normalizing to the dominant line from the upper level, a signal-to-noise ratio (S/N) weighted mean over all spectra yields the BR for that line. BR uncertainties are determined from the standard deviation of the weighted mean, the inverse of the S/Ns and the systematic uncertainty of the calibration. The BRs from the 3-m echelle spectra, for transition wavelengths ranging from 2350 to 4360 Å, are then combined with those from the FTS, wavelengths ranging from 2900 Å to the near-infrared. Bridge lines that appear in both analyses are used to put the two sets of BRs on the same scale, and then the final set of BRs is renormalized to yield BFs. The BF uncertainties of the echelle-only lines is that of the echelle

BRs, while that of the bridge lines and the FTS-only optical lines have an uncertainty of the bridge rescaling added in quadrature with the BR uncertainty. Finally, the FTS-only lines in some cases have an additional systematic calibration uncertainty added in quadrature if there is a significant branch in the far-UV that was not included in the FTS analysis.

*2.2 Blends*

Blends are always a potential problem when measuring BFs, and much more so in the lanthanides because of the richness and complexity of the spectra resulting from many low-lying configurations of opposite parity that arise from open *f*, *d*, *p* and *s*-shells. In the spectrum of a HCL, blends may arise from transitions of Tm II, Tm I, and the first and second spectra of the buffer gas, either Ar or Ne. This, coupled with the presence of Fe I and Fe II contaminant lines in our spectra, means that the possibility of blends is high.

As mentioned above, the analysis software automatically indicates the position of potential blends based on all dipole allowed transitions. Sometimes a blend is obvious based on a broadened line shape, but this is not always the case. A potential blend is confirmed by observing the BR of the line over a range of lamp operating conditions. In past studies, a variety of techniques have been used to separate blends. Blends with the buffer gas species are usually straightforward to resolve simply by evaluating the BR based on spectra in the other buffer gas. The center-of-gravity technique described in Den Hartog, Lawler & Roederer (2021) is sometimes used to separate blends in FTS spectra. It cannot be used in Tm II, however, because the energy levels are not known to high enough accuracy and precision. The technique, in which the center-of-gravity of the blended line is compared to the Ritz wavenumbers of the line of interest and the blending partner to determine the blend fraction, requires better than 1 part in $10^7$ wavenumber accuracy characteristic of the FTS and energy levels known to ~0.001 cm$^{-1}$.

In this study we use a technique described in Den Hartog et al. (2019) that compares the measured BRs of the blended line with clean, unblended lines from the level of interest and from the blending partner upper level over all spectra to solve for the blend fraction in each spectrum. This is an exact solution when comparing a single pair of ratios, or can be solved with a least-squares analysis when comparing to multiple clean lines from either or both upper levels involved in the blend. This technique works to the degree that the two levels involved in the blend have different population dependencies over the range of operating conditions for the spectra evaluated. Experience shows that having high quality spectra in both Ar and Ne buffer gas, such as those used in the echelle analysis, yields the highest range of blend fractions and therefore the most convincing blend separation from the least-squares technique. That being said, we were able to separate a number of blends in the Tm-Ar FTS data by relying only on the current dependence of the blend fraction. Transitions that had a blend separation done within this study are followed by a "d" superscript in the wavelength column of Table 3. The BF uncertainties for these transitions were increased by an estimate of the systematic uncertainty associated with the blend separation.

*2.3 A Note Regarding Hyperfine Structure*

Thulium has only one stable isotope, $^{169}$Tm, with nuclear spin $I = ½$ and a weak magnetic dipole moment of -0.2310(15) nm (Stone 2005). A nuclear spin of ½ results in a simple hyperfine structure (HFS) pattern with two strong components ($\Delta F = \Delta J$) and then one or two much weaker components ($\Delta F = \Delta J \pm 1$). We observe some lines where the two strong HFS components are resolved in our FTS spectra and many more that are partially resolved or noticeably broadened in both the FTS and echelle spectra. The FTS has higher resolving power than the echelle spectrograph. Linewidths in the FTS spectra range upward from a Doppler limited minimum of ~0.08 cm$^{-1}$ for an isolated HFS component to ~0.24 cm$^{-1}$ for the FWHM of a fully resolved HFS pattern. Linewidths in the echelle spectra have additional instrumental broadening and are ~0.13 cm$^{-1}$ wide at a minimum. The amount of instrumental broadening varies depending on the position of the line on the CCD. The HFS does not affect our BF analysis, since we use numerical integrals across the line profiles rather than line fitting to determine our intensities. While it is important to include HFS in astrophysical spectral syntheses for any lines where the HFS is a significant fraction of the Doppler broadened linewidth, an analysis of HFS is beyond the scope of the current project. Fortunately, a recent study has been published by Kebapci et al. (2024) which significantly expands (and corrects) the precise but limited Tm II HFS analysis by Mansour et al. (1989). We recommend this work for HFS magnetic dipole *A* constants for Tm II.

## 3. RESULTS AND DISCUSSION

Our measured BFs are presented in Table 3 for transitions associated with 13 odd-parity and 24 even-parity upper levels. Two of these levels were included in our group's earlier study of Tm II by UW97, whose results are also given in Table 3. The odd-parity level at 39638.41 cm$^{-1}$ is analyzed strictly from the FTS spectra in both studies. It is included in the current study for the purpose of comparing our FTS analyses, and particularly our determination of the instrument sensitivity for each spectrum, with the earlier study. This was deemed prudent due to changes in the software and group personnel in the intervening 25 years. The even-parity level at 33398.70 cm$^{-1}$ was analyzed strictly from the 3-m echelle data in this study, whereas the UW97 study was strictly from analysis of FTS data. This level was chosen for re-measurement to demonstrate that the radiometric calibrations of the echelle and the FTS yield consistent BFs. Examination of Table 3 for these two levels shows excellent agreement between the old study and the current results that are well within the stated uncertainties. This comparison is also made in Figure 1 which shows the difference in the log of the BFs versus the log(BF) from this study. The comparison to UW97 appears in both top and bottom panels (black stars) where the lower panel has a factor-of-10 more sensitive vertical scale in the logarithm. The heavy horizontal line at 0.0 represents perfect agreement.

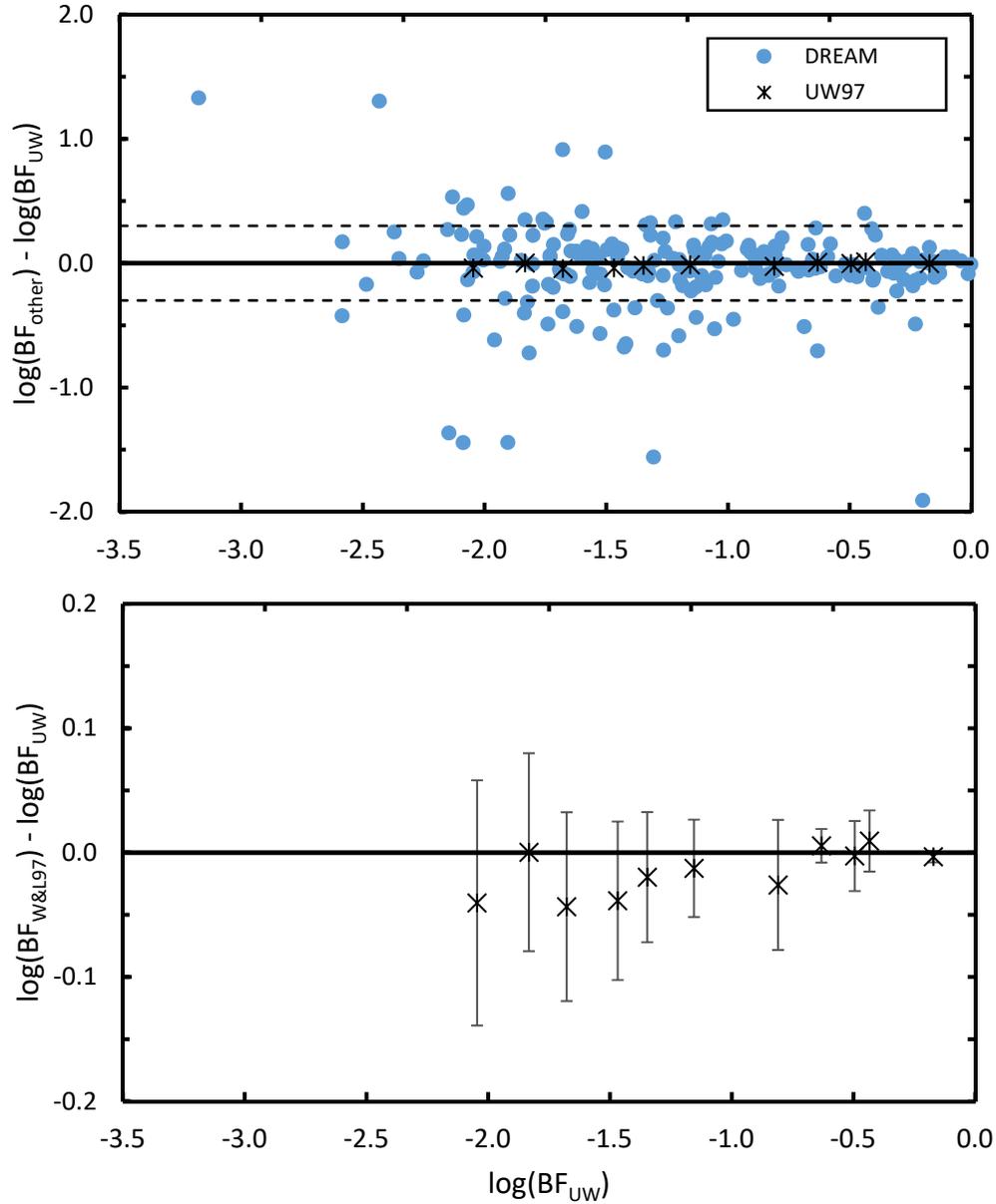

Figure 1. Logarithmic differences of BFs measured in this study to those in the literature. The top panel shows a comparison to BFs calculated from the theoretical *gA*-values of Quinet, Palmeri, & Biémont (1999) downloaded from the D.R.E.A.M. database (blue circles; https://agif.umons.ac.be/databases/dream.html) and to the experimental BFs of UW97 (black stars). The lower panel shows the same comparison with UW97 on an expanded scale. Error bars in the lower panel indicate the uncertainties of UW97 and this study combined in quadrature. The solid horizontal lines at 0.0 in each panel indicate perfect agreement, while the dashed lines in the top panel indicate factor of two difference.

Also shown in the top panel of Figure 1 is a comparison of all our BFs to theoretical results from the Relativistic Hartree-Fock calculations (including configuration interaction and core-polarization effects) of Quinet, Palmeri, & Biémont (1999) (blue circles) downloaded from the D.R.E.A.M.[4] database (Quinet & Palmeri 2020). The significant scatter in this figure reflects the difficulties involved in theoretical calculations of the complicated atomic structure in the lanthanides. With their high-Z nuclei, and configurations including open *f*, *d*, *p*, and *s*-shells, it is a very difficult challenge to calculate accurate transition data, particularly for the weaker branches. And yet, on average, the BFs calculated from their *gA*-values agree with our results within 0.04 dex (~2%) with a standard deviation of 0.37 dex (a factor of ~2.4) improving to 0.20 dex for stronger branches (BF > 0.1). The other theoretical work of Radžiūtė et al. (2021) proved too difficult to compare to, as it is unclear which of their 1100 levels of Tm II correspond to the levels involved in this study.

Our measured BFs are converted to *A*-values and log(*gf*)s following the relations in Martin, et al. (2023)

$$A_{ul} = \frac{BF_{ul}}{\tau_u} \quad ; \quad \log(gf) = \log\left(\frac{1.4992 g_u A_{ul}}{\sigma^2}\right) \quad , \tag{2}$$

where $A_{ul}$ is the transition probability in s$^{-1}$, $\tau_u$ is the radiative lifetime of the upper level in s, $g_u$ is the degeneracy of the upper level, and $\sigma$ is the transition wavenumber in cm$^{-1}$. The radiative lifetimes of UW96 are used to establish the absolute scale for our BFs. The uncertainty of the *A*-value is the uncertainty of the BF and that of the lifetime added in quadrature. We present *A*-values with their uncertainties and log(*gf*)s in Table 4.

## 4. ABUNDANCE DETERMINATION IN AN R-PROCESS-ENHANCED METAL-POOR STAR

One potential application of these new data is to enable the use of additional Tm II lines as abundance indicators in stars. Tm is expected to have the second-lowest abundance (after Ta) of any element with one or more stable isotopes in the atmospheres of metal-poor r-process-enhanced stars, and often only a handful of lines—typically ≤ 6—are useful. We search for the lines listed in Table 4 in the optical and UV spectra of metal-poor star HD 222925 (Roederer et al. 2018, 2022). This star exhibits a high enhancement of r-process elements ([Eu/Fe] = +1.32 ± 0.08) relative to its moderately low metallicity ([Fe/H] = −1.46 ± 0.10). High-resolution optical spectra (3330 – 9410 Å; Roederer et al. 2018) from the Magellan Inamori Kyocera Echelle (MIKE) spectrograph on the Magellan II Telescope at Las Campanas Observatory, Chile, and UV spectra (1936 – 3145 Å; Roederer et al. 2022) from the Space Telescope Imaging Spectrograph (STIS) on board the Hubble Space Telescope (HST) are available for HD 222925.

---

[4] Available online at: https://agif.umons.ac.be/databases/dream.html. The database does not directly report BFs. We have calculated BFs from their *gA*-values.

We follow the procedure outlined in Den Hartog et al. (2021) to synthesize the stellar spectrum around the 25 potentially strongest Tm II lines in HD 222925. This procedure can be briefly summarized as follows. The relative strength for each line is calculated following Sneden et al. (2009). A list of all potentially viable transitions within 3 Å of each line of interest is generated using the LINEMAKE code (Placco et al. 2021). The ATLAS9 (Castelli & Kurucz 2004) model atmosphere used by Roederer et al. (2018) is also used here (effective temperature 5636 ± 103 K, log of surface gravity 2.54 ± 0.17, microturbulent velocity parameter 2.20 ± 0.20 km s$^{-1}$, model metallicity −1.5 ± 0.1). A recent version of the MOOG line analysis software (Sneden 1973; Sobeck et al. 2011; 2017 version[5]) is used to synthesize these spectral lines and compare the synthetic spectra to the observed ones.

Among the ≈5 – 10 Tm II lines that are covered by these spectra and are potentially strong enough to be detectable, only one is not blended with other spectral features. This line, at 2624.34 Å, is shown in Figure 2. The overall continuum is depressed by ≈15 – 20% at this wavelength, due to the presence of a strong Fe II line at 2625.67 Å, but the spectral region can be reasonably well fit. We considered whether HFS needed to be included in the spectral fitting of this line and decided it did not. The hyperfine $A$'s from Kebapci (2024) for the upper and lower levels yield a HFS splitting of ~0.005 Å for the two strong components of this transition, which is a very small fraction of the Doppler width of the absorption feature shown in Figure 2. This Tm II line yields an abundance of log ε = −0.05 ± 0.20, which is in good agreement with the mean Tm abundance derived from 10 other ultraviolet and optical lines of Tm II, log ε = −0.09 ± 0.10 (Roederer et al. 2022).

We also checked for relatively strong Tm II lines in the optical spectrum of the r-process-enhanced metal-poor star 2MASS J22132050−5137385 (Roederer et al. 2024) and the optical and UV spectra of the s-process-enhanced star HD 196944 (Placco et al. 2015; see data references in Den Hartog et al. 2021). No previously unstudied Tm II lines from Table 4 are detectable in either star. Even though our analysis has revealed only one new Tm abundance indicator in HD 222925, it remains important to identify any potentially useful lines when the number of such is intrinsically small due to the low elemental abundance. Furthermore, these new data have been incorporated into the LINEMAKE database, where they may also be useful when modeling Tm II lines that blend with other lines of interest in stellar spectra.

---

[5] http://www.github.com/alexji/moog17scat

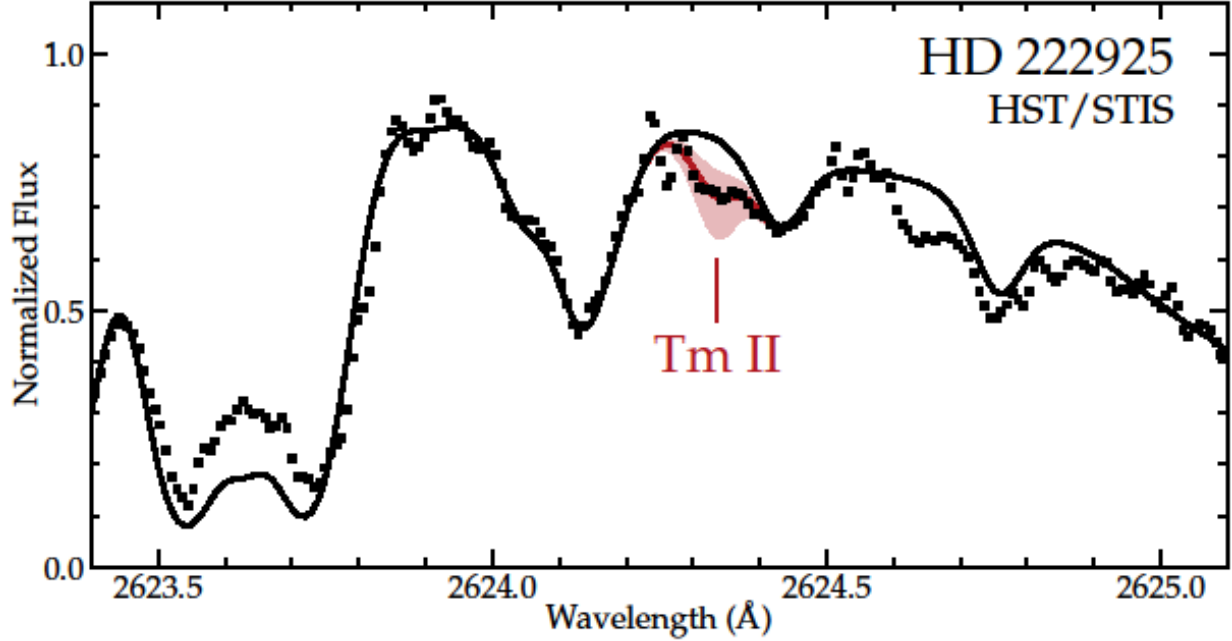

Figure 2: Comparison of synthetic (lines) and observed (points) spectra around the Tm II line at 2624.34 Å in HD 222925. The red line shows the best-fit abundance, and the shaded regions indicate variations in this best-fit abundance by factors of ± 2 (i.e., 0.30 dex). The black line shows a spectrum with no Tm.

## 5. SUMMARY

We report new experimental BF measurements for 224 UV and optical transitions of Tm II associated with 37 upper levels. Thirty-five of the levels including 213 transitions are high-lying levels studied for the first time. The BFs are combined with radiative lifetimes from an earlier study to determine $A$-values and log($gf$)s. The uncertainties of the final $A$-values range from 5% on the strongest branches to a maximum of 30% on the weakest lines reported. Comparison is made to earlier experimental and theoretical studies from the literature. These new data are employed to identify previously unused Tm II lines in metal-poor stars that could potentially be used as abundance indicators. One new Tm II line, at 2624.34 Å, yields an abundance consistent with previous determinations based on other UV and optical Tm II lines.

## ACKNOWLEDGMENTS

E.A.D.H. acknowledges support from NSF grant AST 2206050. I.U.R. acknowledges support from grants GO-15657 and GO-17166 from the Space Telescope Science Institute, which is operated by the Association of Universities for Research in Astronomy, Incorporated, under NASA contract NAS5-26555, and from the U.S. National Science Foundation (NSF): grants PHY 14-30152 (Physics Frontier Center/JINA-CEE) and AST 2205847.

Figure Captions:

Figure 1: Logarithmic differences of BFs measured in this study to those in the literature. The top panel shows a comparison to BFs calculated from the theoretical *gA*-values of Quinet, Palmeri, & Biémont (1999) downloaded from the D.R.E.A.M. database (blue circles; htps://agif.umons.ac.be/databases/dream.html) and to the experimental BFs of UW97 (black stars). The lower panel shows the same comparison with UW97 on an expanded scale. Error bars in the lower panel indicate the uncertainties of UW97 and this study combined in quadrature. The solid horizontal lines at 0.0 in each panel indicate perfect agreement, while the dashed lines in the top panel indicate factor-of-two difference.

Figure 2: Comparison of synthetic (lines) and observed (points) spectra around the Tm II line at 2624.34 Å in HD 222925. The red line shows the best-fit abundance, and the shaded regions indicate variations in this best-fit abundance by factors of ± 2 (i.e., 0.30 dex). The black line shows a spectrum with no Tm.

# REFERENCES

Anderson, H. M., Den Hartog, E. A. & Lawler, J. E. 1996, J. Opt. Soc. Am. B 13, 2382
    DOI:10.1364/JOSAB.13.002382

Castelli, F., & Kurucz, R. L. 2004, arXiv:astro-ph/0405087

Cowan, J. J., Sneden, C., Lawler, J. E., et al. 2021, Rev. Mod. Phys. 93, 0151002

Danzmann, K., & Kock, M. 1982, J. Opt. Soc. Am., 72, 1556

Den Hartog, E. A., Lawler, J. E., & Roederer, I. U. 2021, ApJS 254:5 DOI: 10.3847/1538-4365/abe861

Den Hartog, E. A., Lawler, J. E., Sneden, C, Cowan, J. J., & Brukhovesky, A. 2019, ApJS 243:33 DOI: 10.3847/1538-4365/ab322e

Den Hartog, E. A., Lawler, J. E., Sneden, C, Roederer, I. U. & Cowan, J. J. 2023, ApJS 265:42. DOI:10.3847/1538-4365/acb642

Drout, M. R., Piro, A. L., Shappee, B. J., et al. 2017, Science 358, 1570

Frischknecht, U., Hirschi, R., Pignatari, M., et al. 2016, MNRAS 456, 1803

Gull, M., Frebel, A., Hinojosa, K., et al. 2021, ApJ 912, 52

Hashiguchi, S., & Hasikuni, M. 1985, J. Phys. Soc. Japan, 54, 1290

Honda, S., Aoki, W., Ishimaru, Y., Wanajo, S., Ryan, S. G. 2006, ApJ 643, 1180

Karakas, A. I. 2010, MNRAS 403, 1413

Kasen, D., Badnell, N. R., Barnes, J. 2013, ApJ 774, 25

Kebapci, T. Y., Parlatan, Ş., Sert, S. et al. 2024, arXiv:2405.05309v1 [astro-ph.IM]

Kramida, A., Ralchenko, Yu., Reader, J., and NIST ASD Team (2022). NIST Atomic Spectra Database (ver. 5.10), [Online]. Available: https://physics.nist.gov/asd [2023, September 26]. National Institute of Standards and Technology, Gaithersburg, MD. DOI:10.18434/T4W30F

Lawler, J. E., Sneden, C., Cowan, J. J., Ivans, I. I., & Den Hartog, E. A. 2009, ApJS, 182, 51
    DOI:10.1088/0067-0049/182/1/51

Magrini, L., Spina, L., Randich, S., et al. 2018, A&A 617, A106

Mansour, N. B., Dinneen, T.P. & Young, L. 1989, NIMPR B40/41, 252

Martin, W. C., Wiese, W. L. & Kramida, A., *Atomic Spectroscopy,* Ch. 11 in Springer Handbook of Atomic, Molecular, and Optical Physics (Edited by G. W. F. Drake, Springer, Cham, 2023). doi:10.1007/978-3-030-73893-8_11

Martin, W. C., Zalubas, R. & Hagan, L. 1978, in Nat. Stand. Ref. Data Ser., NSRDS-NBS 60, 422 pp. (Nat. Bur. Stand., U.S., 1978) DOI:10.6028/NBS.NSRDS.60

Moore, C. E., Minnaert, M. G. J., & Houtgast, J. 1966, The Solar Spectrum 2935 A to 8770 Å (Nat. Bur. Stds. Monograph; Washington, DC: US Govt. Printing Office)

Peck, E. R., & Reeder, K. 1972, JOSA, 62, 958

Placco, V. M., Beers, T. C., Ivans, I. I., et al. 2015, ApJ 812, 109

Placco, V. M., Sneden, C., Roederer, I. U., et al. 2021, RNAAS 5, 92

Quinet P. & Palmeri P. 2020, Atoms 8, 18

Quinet, P. Palmeri, P. & Biémont, E. 1999, J. Quant. Spectrosc. Radiat. Transfer 62, 625
    DOI:10.1016/S0022-4073(98)00127-7
Radžiūtė, L., Gaigalas, G., Kato, D., Rynkun, P. & Tanaka, M. 2021 Astrophys. J., Suppl. Ser. 257, 29 DOI:10.3847/1538-4365/ac1ad2
Rieger, G., McCurdy, M. M., & Pinnington, E. H. 1999, Phys. Rev. A 60, 4150
    DOI:10.1103/PhysRevA.60.4150
Roederer, I. U., Beers, T. C., Hattori, K., et al. 2024, ApJ, submitted
Roederer, I. U., Lawler, J. E., Den Hartog, E. A., et al. 2022, ApJS 260, 27
Roederer, I. U., Sakari, C. M., Placco, V. M., et al. 2018, ApJ 865, 129
Roederer, I. U., Vassh, N., Holmbeck, E. M., et al. 2023, Science 382, 1177
Simmerer, J., Sneden, C., Cowan, J. J., et al. 2004, ApJ 617, 1091
Sneden, C. A. 1973, PhD thesis, Univ. Texas at Austin
Sneden, C., Cowan, J. J., Gallino, R. 2008, ARAA 46, 241
Sneden, C., Lawler, J. E., Cowan, J. J., Ivans, I. I., & Den Hartog, E. A. 2009, ApJS 182, 80
Sobeck, J. S., Kraft, R. P., Sneden, C., et al. 2011, AJ 141, 175
Stone, N. J. 2005, ADNDT 90, 75
Tanvir, N. R., Levan, A. J., González-Fernández, C., et al. 2017, ApJL 848, L27
Tian, Y.-S. Wang, X.-H., Yu, Qi, Li, Y.-F., Gao, Y. & Dai, Z.-W. 2016, Mon. Not. R. Astron. Soc. 457, 1393 DOI:10.1093/mnras/stv3015
Wang, H. S., Morel, T., Quanz, S. P., Mojzsis, S. J. 2020, A&A 644, A19
Wang, X.-H., Yu, Qi, Tian, Y.-S., Chen, Z.-M., Xie, H.-Q., Zeng, X.-Y., Guo, G.-C. & Chang, H.-X. 2022, J. Quant. Spectrosc. Radiat. Transfer 280, 108091
    DOI:10.1016/j.jqsrt.2022.108091
Whaling, W., Carle, M. T., & Pitt, M. L. 1993, J. Quant. Spectrosc. Radiat. Transfer, 50, 7
Wickliffe, M. E., & Lawler, J. E. 1997, J. Opt. Soc. Am. B, 14, 737
Wood, M. P. & Lawler, J. E. 2012, ApOpt, 51, 8407 DOI:10.3847/1538-4365/ab84f5
Wyart, J.-F. 2011, Can. J. Phys. 89, 451 DOI:10.1139/p10-112
Xu, H.-L., Jiang, Z.-K. & Svanberg, S. 2003, Eur. Phys. J. D 23, 323
    DOI:10.1140/epjd/e2003-00055-3

Table 1
FTS spectra[a] of Tm HCLs[b]

| Index | Date | Serial Num. | Buffer Gas | Lamp Current (mA) | Wavenumber Range (cm$^{-1}$) | # scan | Beam-Splitter | Filter | Diode Detector |
|---|---|---|---|---|---|---|---|---|---|
| 1 | 1996 Apr 12 | 25 | Ar | 18.8 | 8438 – 30003 | 86 | visible | none | superblue |
| 2 | 1995 Mar 28 | 1 | Ar | 19.5 | 14865 – 35071 | 8 | UV | CuSO$_4$ | midrange |
| 4 | 1995 Mar 28 | 3 | Ar | 19.5 | 8371 – 35071 | 8 | UV | none | midrange |
| 6 | 1995 Mar 28 | 5 | Ar | 24.8 | 8371 – 35071 | 8 | UV | none | midrange |
| 7 | 1995 Mar 29 | 17 | Ar | 22.9 | 8371 – 35071 | 8 | UV | none | midrange |
| 8 | 1995 Mar 29 | 18 | Ar | 22.7 | 8371 – 35071 | 8 | UV | none | midrange |
| 9 | 1995 Mar 30 | 19 | Ar | 38.0 | 8371 – 35071 | 8 | UV | none | midrange |
| 10 | 1996 Apr 11 | 20 | Ar | 214 | 8438 – 30003 | 8 | visible | none | superblue |
| 11 | 1996 Apr 10 | 9 | Ar | 220 | 8438 – 30003 | 2 | visible | none | superblue |
| 13 | 1995 Mar 28 | 6 | Ne | 18.3 | 8371 – 35071 | 8 | UV | none | midrange |
| 14 | 1995 Mar 28 | 7 | Ne | 24.5 | 8371 – 35071 | 8 | UV | none | midrange |

Notes:
[a] All spectra were recorded using the 1 m FTS on the McMath-Pierce Solar telescope at the National Solar Observatory, Kitt Peak, AZ. FTS was operated with spectral resolution of 0.053 cm$^{-1}$ for all spectra.
[b] Lamps used were commercially available HCLs with fused-silica windows and forced-air cooling except for spectra with indices 10 and 11 which utilized a water-cooled demountable HCL with Suprasil 2 window

Table 2. Echelle spectra of commercial HCLs used in the study of Tm II BFs.[a]

| Index[b] | Date | Serial Number | Buffer Gas | Lamp Current (mA) | Frame Desig.[c] | Accums | Total Exposure (min) |
|---|---|---|---|---|---|---|---|
| 21 | 2023 Feb 25 | 1 | Neon | 10 | C | 160 | 20 |
| 22 | 2023 Feb 25 | 3 | Neon | 10 | B | 240 | 30 |
| 23 | 2023 Feb 25 | 5 | Neon | 10 | C | 240 | 30 |
| 24 | 2023 Feb 25 | 7 | Neon | 10 | D | 108 | 31 |
| 25 | 2023 Feb 25 | 9 | Neon | 10 | C | 240 | 30 |
| 26 | 2023 Mar 11 | 1 | Neon | 15 | C | 300 | 15 |
| 27 | 2023 Mar 11 | 3 | Neon | 15 | B | 600 | 30 |
| 28 | 2023 Mar 11 | 5 | Neon | 15 | C | 600 | 30 |
| 29 | 2023 Mar 11 | 7 | Neon | 15 | D | 330 | 30 |
| 30 | 2023 Apr 1 | 1 | Neon | 20 | B | 705 | 20 |
| 31 | 2023 Apr 1 | 3 | Neon | 20 | C | 705 | 20 |
| 32 | 2023 Apr 1 | 5 | Neon | 20 | D | 480 | 20 |
| 33 | 2023 Apr 1 | 7 | Neon | 20 | B | 705 | 20 |
| 34 | 2023 Apr 8 | 1 | Neon | 20 | C | 343 | 60 |
| 35 | 2023 Apr 15 | 1 | Neon | 10 | C | 105 | 70 |
| 36 | 2023 Apr 15 | 3 | Neon | 10 | B | 105 | 70 |
| 37 | 2023 Apr 15 | 5 | Neon | 10 | C | 135 | 90 |
| 38 | 2023 Apr 15 | 7 | Neon | 10 | D | 105 | 70 |
| 39 | 2023 Apr 22 | 1 | Neon | 15 | B | 90 | 60 |
| 40 | 2023 Apr 22 | 3 | Neon | 15 | C | 144 | 60 |
| 41 | 2023 Apr 22 | 5 | Neon | 15 | B | 120 | 80 |
| 42 | 2023 Apr 22 | 7 | Neon | 15 | D | 90 | 83 |
| 43 | 2023 Jun 8 | 1 | Neon | 20 | C | 270 | 90 |
| 44 | 2023 Jun 8 | 3 | Neon | 20 | B | 180 | 90 |
| 45 | 2023 Jun 8 | 5 | Neon | 20 | C | 270 | 90 |
| 46 | 2023 Jun 8 | 7 | Neon | 20 | D | 193 | 90 |
| 47 | 2023 Jun 13 | 1 | Neon | 25 | C | 257 | 60 |
| 48 | 2023 Jun 13 | 3 | Neon | 25 | B | 208 | 90 |
| 49 | 2023 Jun 13 | 5 | Neon | 25 | C | 386 | 90 |
| 50 | 2023 Jun 13 | 7 | Neon | 25 | D | 235 | 90 |
| 51 | 2023 Jun 19 | 1 | Argon | 12 | C | 360 | 90 |
| 52 | 2023 Jun 19 | 3 | Argon | 12 | B | 225 | 90 |
| 53 | 2023 Jun 19 | 5 | Argon | 12 | C | 360 | 90 |
| 54 | 2023 Jun 19 | 7 | Argon | 12 | D | 257 | 90 |
| 55 | 2023 Jun 21 | 1 | Argon | 20 | C | 360 | 30 |
| 56 | 2023 Jun 21 | 3 | Argon | 20 | B | 831 | 90 |
| 57 | 2023 Jun 21 | 5 | Argon | 20 | C | 900 | 75 |
| 58 | 2023 Jun 21 | 7 | Argon | 20 | D | 831 | 90 |

Note:
[a] All echelle spectra were taken from commercially manufactured Tm-Ne or Tm-Ar HCLs and have a spectral coverage from 2350 – 4400 Å in the low-resolution direction and resolving power of ~250,000, although the effective resolving power is somewhat lower due to line broadening. Each HCL spectrum was calibrated with a $D_2$ lamp spectrum, which was recorded immediately following the completion of the HCL spectrum.
[b] Each spectrum listed is a single CCD frame and does not cover an entire echelle grating order. A minimum of three overlapping frames are needed per grating order in the UV. Four or five are recorded for each data set to give some redundancy.
[c] Frame Designation of C indicates the CCD frame straddles the center of the grating order, B is shifted toward lower wavelengths and D toward higher wavelengths. The breadth of B-C-D frames covers one grating order in the UV.

Table 3
BFs of Tm II

| Upper Level[a] | | Lower Level[a] | | $\lambda_{air}$ | $\sigma_{vac}$ | This Study | | UW97[b] | |
|---|---|---|---|---|---|---|---|---|---|
| $E_k$ (cm$^{-1}$) | $J_k$ | $E_i$ (cm$^{-1}$) | $J_i$ | (Å) | (cm$^{-1}$) | BF[c] | unc. (%) | BF | unc. (%) |
| | | | | Odd-parity Upper Levels | | | | | |
| 39638.41 | 7 | 12457.29 | 6 | 3677.98 | 27181.12 | 0.368 | 5 | 0.376 | 3 |
| | | 19526.82 | 6 | 4970.87 | 20111.59 | 0.070 | 5 | 0.068 | 8 |
| | | 19619.08 | 7 | 4993.78 | 20019.33 | 0.045 | 8 | 0.043 | 10 |
| | | 19682.97 | 8 | 5009.77 | 19955.44 | 0.320 | 3 | 0.318 | 6 |
| | | 22052.39 | 7 | 5684.76 | 17586.02 | 0.155 | 8 | 0.146 | 10 |
| | | 22355.43 | 6 | 5784.44 | 17282.98 | 0.021 | 14 | 0.019 | 13 |
| | | 23961.60 | 7 | 6377.09 | 15676.81 | 0.009 | 17 | 0.008 | 19 |
| | | 24059.08 | 8 | 6416.99 | 15579.33 | 0.011 | 28 | … | … |
| | | | | | residual | 0.001 | | 0.022 | |
| 47808.54 | 5 | 12457.29 | 6 | 2827.92 | 35351.25 | 0.589 | 2 | … | … |
| | | 16567.47 | 5 | 3199.99 | 31241.07 | 0.080 | 5 | … | … |
| | | 19526.82 | 6 | 3534.84 | 28281.72 | 0.264 | 5 | … | … |
| | | 20619.05 | 5 | 3676.85 | 27189.49 | 0.0026 | 20 | … | … |
| | | 21021.98 | 5 | 3732.15 | 26786.56 | 0.029 | 9 | … | … |
| | | 23904.43 | 6 | 4182.20 | 23904.11 | 0.0033 | 19 | … | … |
| | | 30508.76 | 4 | 5778.82 | 17299.78 | 0.016 | 29 | … | … |
| | | | | | residual | 0.016 | | | |
| 47927.14 | 6 | 12457.29 | 6 | 2818.47 | 35469.85 | 0.529 | 3 | … | … |
| | | 19526.82 | 6 | 3520.08 | 28400.32 | 0.065 | 12 | … | … |
| | | 20619.05 | 5 | 3660.88 | 27308.09 | 0.156 | 7 | … | … |
| | | 21021.98 | 5 | 3715.70 | 26905.16 | 0.030 | 11 | … | … |
| | | 23904.43 | 6 | 4161.55 | 24022.71 | 0.0085 | 12 | … | … |
| | | 26256.26 | 7 | 4613.20 | 21670.88 | 0.047 | 20 | … | … |
| | | 26709.43 | 6 | 4711.73 | 21217.71 | 0.038 | 20 | … | … |
| | | 28095.31 | 7 | 5040.99[d] | 19831.83 | 0.061 | 16 | … | … |
| | | 30361.06 | 6 | 5691.21 | 17566.08 | 0.060 | 20 | … | … |
| | | | | | residual | 0.006 | | | |
| 48195.94 | 6 | 12457.29 | 6 | 2797.27 | 35738.65 | 0.617 | 2 | … | … |
| | | 16567.47 | 5 | 3160.79 | 31628.47 | 0.0074 | 9 | … | … |
| | | 19526.82 | 6 | 3487.08 | 28669.12 | 0.122 | 8 | … | … |
| | | 21021.98 | 5 | 3678.95 | 27173.96 | 0.160 | 9 | … | … |
| | | 22052.39 | 7 | 3823.95[d] | 26143.55 | 0.009 | 14 | … | … |
| | | 25014.75 | 5 | 4312.63 | 23181.19 | 0.010 | 14 | … | … |
| | | 26256.26 | 7 | 4556.67 | 21939.68 | 0.037 | 15 | … | … |
| | | 28095.31 | 7 | 4973.58 | 20100.63 | 0.018 | 20 | … | … |
| | | | | | residual | 0.019 | | | |

| Upper Level[a] | | Lower Level[a] | | $\lambda_{\text{air}}$ | $\sigma_{\text{vac}}$ | This Study | | UW97[b] | |
|---|---|---|---|---|---|---|---|---|---|
| $E_k$ (cm$^{-1}$) | $J_k$ | $E_i$ (cm$^{-1}$) | $J_i$ | (Å) | (cm$^{-1}$) | BF[c] | unc. (%) | BF | unc. (%) |
| 49765.68 | 7 | 12457.29 | 6 | 2679.57 | 37308.39 | 0.319 | 3 | ... | ... |
| | | 19619.08 | 7 | 3316.17 | 30146.60 | 0.081 | 8 | ... | ... |
| | | 19682.97 | 8 | 3323.21 | 30082.71 | 0.136 | 7 | ... | ... |
| | | 22052.39 | 7 | 3607.35 | 27713.29 | 0.141 | 9 | ... | ... |
| | | 22355.43 | 6 | 3647.23[d] | 27410.25 | 0.167 | 9 | ... | ... |
| | | 23024.29 | 8 | 3738.46 | 26741.39 | 0.016 | 13 | ... | ... |
| | | 23904.43 | 6 | 3865.69 | 25861.25 | 0.015 | 12 | ... | ... |
| | | 24059.08 | 8 | 3888.95[d] | 25706.60 | 0.047 | 12 | ... | ... |
| | | 26256.26 | 7 | 4252.42 | 23509.42 | 0.028 | 14 | ... | ... |
| | | 26478.28 | 6 | 4292.96 | 23287.40 | 0.026 | 15 | ... | ... |
| | | 26709.43 | 6 | 4336.00 | 23056.25 | 0.015 | 15 | ... | ... |
| | | | | | residual | 0.010 | | | |
| 50617.56 | 6 | 16567.47 | 5 | 2935.99 | 34050.09 | 0.744 | 2 | ... | ... |
| | | 19526.82 | 6 | 3215.46 | 31090.74 | 0.019 | 9 | ... | ... |
| | | 21021.98 | 5 | 3377.91 | 29595.58 | 0.050 | 5 | ... | ... |
| | | 22052.39 | 7 | 3499.77 | 28565.17 | 0.027 | 7 | ... | ... |
| | | 22355.43 | 6 | 3537.29 | 28262.13 | 0.012 | 10 | ... | ... |
| | | 23768.84 | 5 | 3723.51 | 26848.72 | 0.005 | 25 | ... | ... |
| | | 25014.75 | 5 | 3904.72 | 25602.81 | 0.021 | 12 | ... | ... |
| | | 26256.26 | 7 | 4103.71 | 24361.30 | 0.035 | 11 | ... | ... |
| | | 33182.00 | 7 | 5733.82 | 17435.56 | 0.044 | 19 | ... | ... |
| | | | | | residual | 0.042 | | | |
| 51116.29 | 8 | 19619.08 | 7 | 3173.97 | 31497.21 | 0.019 | 6 | ... | ... |
| | | 19682.97 | 8 | 3180.42 | 31433.32 | 0.055 | 3 | ... | ... |
| | | 20465.82 | 9 | 3261.65 | 30650.47 | 0.513 | 1 | ... | ... |
| | | 22052.39 | 7 | 3439.71 | 29063.90 | 0.020 | 5 | ... | ... |
| | | 23024.29 | 8 | 3558.72 | 28092.00 | 0.072 | 4 | ... | ... |
| | | 24059.08 | 8 | 3694.82[d] | 27057.21 | 0.240 | 7 | ... | ... |
| | | 26256.26 | 7 | 4021.39 | 24860.03 | 0.043 | 6 | ... | ... |
| | | 33182.00 | 7 | 5574.36 | 17934.29 | 0.023 | 19 | ... | ... |
| | | 34027.94 | 8 | 5850.32 | 17088.35 | 0.015 | 30 | ... | ... |
| | | | | | residual | 0.000 | | | |
| 52331.33 | 7 | 12457.29 | 6 | 2507.14 | 39874.04 | 0.091 | 7 | ... | ... |
| | | 19526.82 | 6 | 3047.48 | 32804.51 | 0.050 | 5 | ... | ... |
| | | 19619.08 | 7 | 3056.07 | 32712.25 | 0.480 | 2 | ... | ... |
| | | 19682.97 | 8 | 3062.05 | 32648.36 | 0.085 | 2 | ... | ... |
| | | 22052.39 | 7 | 3301.68[d] | 30278.94 | 0.038 | 7 | ... | ... |
| | | 22355.43 | 6 | 3335.05 | 29975.90 | 0.066 | 5 | ... | ... |
| | | 23961.60 | 7 | 3523.88 | 28369.73 | 0.004 | 25 | ... | ... |
| | | 24059.08 | 8 | 3536.03 | 28272.25 | 0.048 | 6 | ... | ... |

| Upper Level[a] | | Lower Level[a] | | $\lambda_{air}$ | $\sigma_{vac}$ | This Study | | UW97[b] | |
|---|---|---|---|---|---|---|---|---|---|
| $E_k$ (cm$^{-1}$) | $J_k$ | $E_i$ (cm$^{-1}$) | $J_i$ | (Å) | (cm$^{-1}$) | BF[c] | unc. (%) | BF | unc. (%) |
| | | 26256.26 | 7 | 3833.99 | 26075.07 | 0.022 | 25 | ... | ... |
| | | 26478.28 | 6 | 3866.92 | 25853.05 | 0.022 | 9 | ... | ... |
| | | 26709.43 | 6 | 3901.81 | 25621.90 | 0.007 | 23 | ... | ... |
| | | 34027.94 | 8 | 5461.95 | 18303.39 | 0.062 | 19 | ... | ... |
| | | 35195.12 | 7 | 5833.98 | 17136.21 | 0.021 | 28 | ... | ... |
| | | | | | residual | 0.004 | | | |
| 53013.80 | 6 | 12457.29 | 6 | 2464.95 | 40556.51 | 0.063 | 7 | ... | ... |
| | | 16567.47 | 5 | 2742.95 | 36446.33 | 0.128 | 4 | ... | ... |
| | | 19526.82 | 6 | 2985.36 | 33486.98 | 0.192 | 2 | ... | ... |
| | | 19619.08 | 7 | 2993.61 | 33394.72 | 0.0082 | 8 | ... | ... |
| | | 21021.98 | 5 | 3124.89 | 31991.82 | 0.147 | 3 | ... | ... |
| | | 22052.39 | 7 | 3228.90 | 30961.41 | 0.146 | 3 | ... | ... |
| | | 22355.43 | 6 | 3260.81 | 30658.37 | 0.013 | 9 | ... | ... |
| | | 23768.84 | 5 | 3418.41 | 29244.96 | 0.073 | 5 | ... | ... |
| | | 23961.60 | 7 | 3441.09 | 29052.20 | 0.063 | 7 | ... | ... |
| | | 26256.26 | 7 | 3736.20 | 26757.54 | 0.041 | 21 | ... | ... |
| | | 26478.28 | 6 | 3767.46 | 26535.52 | 0.026 | 8 | ... | ... |
| | | 33036.56 | 5 | 5004.30[d] | 19977.24 | 0.057 | 30 | ... | ... |
| | | 33182.00 | 7 | 5041.00[d] | 19831.80 | 0.030 | 18 | ... | ... |
| | | | | | residual | 0.013 | | | |
| 53304.88 | 7 | 12457.29 | 6 | 2447.38 | 40847.59 | 0.043 | 7 | ... | ... |
| | | 19526.82 | 6 | 2959.64 | 33778.06 | 0.215 | 6 | ... | ... |
| | | 19619.08 | 7 | 2967.74 | 33685.80 | 0.141 | 3 | ... | ... |
| | | 19682.97 | 8 | 2973.38[d] | 33621.91 | 0.229 | 10 | ... | ... |
| | | 22052.39 | 7 | 3198.82 | 31252.49 | 0.035 | 5 | ... | ... |
| | | 22355.43 | 6 | 3230.14 | 30949.45 | 0.078 | 4 | ... | ... |
| | | 23024.29 | 8 | 3301.50 | 30280.59 | 0.075 | 5 | ... | ... |
| | | 23904.43 | 6 | 3400.33 | 29400.45 | 0.032 | 5 | ... | ... |
| | | 26478.28 | 6 | 3726.58 | 26826.60 | 0.024 | 8 | ... | ... |
| | | 26709.43 | 6 | 3758.97 | 26595.45 | 0.015 | 12 | ... | ... |
| | | 30361.06 | 6 | 4357.25 | 22943.82 | 0.012 | 15 | ... | ... |
| | | | | | residual | 0.101 | | | |
| 53336.89 | 5 | 12457.29 | 6 | 2445.47 | 40879.60 | 0.124 | 5 | ... | ... |
| | | 16567.47 | 5 | 2718.85 | 36769.42 | 0.010 | 18 | ... | ... |
| | | 17974.30 | 4 | 2827.02 | 35362.59 | 0.415 | 2 | ... | ... |
| | | 18291.37 | 4 | 2852.59 | 35045.52 | 0.031 | 4 | ... | ... |
| | | 20619.05 | 5 | 3055.55 | 32717.84 | 0.018 | 6 | ... | ... |
| | | 21021.98 | 5 | 3093.65[d] | 32314.91 | 0.063 | 10 | ... | ... |
| | | 22308.82 | 4 | 3221.96 | 31028.07 | 0.012 | 12 | ... | ... |
| | | 22355.43 | 6 | 3226.81 | 30981.46 | 0.130 | 4 | ... | ... |

| Upper Level[a] | | Lower Level[a] | | $\lambda_{air}$ | $\sigma_{vac}$ | This Study | | UW97[b] | |
|---|---|---|---|---|---|---|---|---|---|
| $E_k$ (cm$^{-1}$) | $J_k$ | $E_i$ (cm$^{-1}$) | $J_i$ | (Å) | (cm$^{-1}$) | BF[c] | unc. (%) | BF | unc. (%) |
| | | 23768.84 | 5 | 3381.06 | 29568.05 | 0.031 | 10 | ... | ... |
| | | 23904.43 | 6 | 3396.63 | 29432.46 | 0.039 | 7 | ... | ... |
| | | 25014.75 | 5 | 3529.80 | 28322.14 | 0.028 | 8 | ... | ... |
| | | 26574.66 | 4 | 3735.55 | 26762.23 | 0.012 | 13 | ... | ... |
| | | 26709.43 | 6 | 3754.45 | 26627.46 | 0.009 | 16 | ... | ... |
| | | 28096.17 | 6 | 3960.73 | 25240.72 | 0.006 | 26 | ... | ... |
| | | | | | residual | 0.073 | | | |
| 53738.88 | 7 | 12457.29 | 6 | 2421.65 | 41281.59 | 0.084 | 11 | ... | ... |
| | | 19526.82 | 6 | 2922.09 | 34212.06 | 0.105 | 5 | ... | ... |
| | | 19619.08 | 7 | 2929.99 | 34119.80 | 0.018 | 10 | ... | ... |
| | | 19682.97 | 8 | 2935.49[d] | 34055.91 | 0.042 | 16 | ... | ... |
| | | 22052.39 | 7 | 3155.01 | 31686.49 | 0.012 | 25 | ... | ... |
| | | 22355.43 | 6 | 3185.47[d] | 31383.45 | 0.086 | 28 | ... | ... |
| | | 23024.29 | 8 | 3254.84 | 30714.59 | 0.074 | 4 | ... | ... |
| | | 23904.43 | 6 | 3350.87 | 29834.45 | 0.033 | 12 | ... | ... |
| | | 23961.60 | 7 | 3357.30[d] | 29777.28 | 0.048 | 10 | ... | ... |
| | | 24059.08 | 8 | 3368.33 | 29679.80 | 0.216 | 3 | ... | ... |
| | | 26256.26 | 7 | 3637.63 | 27482.62 | 0.064 | 4 | ... | ... |
| | | 26478.28 | 6 | 3667.25[d] | 27260.60 | 0.094 | 8 | ... | ... |
| | | 34766.81 | 8 | 5269.44 | 18972.07 | 0.047 | 22 | ... | ... |
| | | | | | residual | 0.077 | | | |
| 54988.96 | 7 | 12457.29 | 6 | 2350.47 | 42531.67 | 0.007 | 27 | ... | ... |
| | | 19526.82 | 6 | 2819.08[d] | 35462.14 | 0.114 | 10 | ... | ... |
| | | 19619.08 | 7 | 2826.43 | 35369.88 | 0.120 | 6 | ... | ... |
| | | 19682.97 | 8 | 2831.55 | 35305.99 | 0.257 | 4 | ... | ... |
| | | 22355.43 | 6 | 3063.44 | 32633.53 | 0.017 | 20 | ... | ... |
| | | 23024.29 | 8 | 3127.55 | 31964.67 | 0.071 | 4 | ... | ... |
| | | 23904.43 | 6 | 3216.11 | 31084.53 | 0.145 | 4 | ... | ... |
| | | 23961.60 | 7 | 3222.03 | 31027.36 | 0.089 | 4 | ... | ... |
| | | 24059.08 | 8 | 3232.19 | 30929.88 | 0.009 | 12 | ... | ... |
| | | 31323.73 | 6 | 4224.42 | 23665.23 | 0.0026 | 18 | ... | ... |
| | | 37104.63 | 8 | 5589.93 | 17884.33 | 0.059 | 22 | ... | ... |
| | | 38537.96 | 6 | 6076.98 | 16451.00 | 0.024 | 31 | ... | ... |
| | | | | | residual | 0.086 | | | |

| Upper Level[a] | | Lower Level[a] | | $\lambda_{air}$ | $\sigma_{vac}$ | This Study | | UW97[b] | |
|---|---|---|---|---|---|---|---|---|---|
| $E_k$ (cm$^{-1}$) | $J_k$ | $E_i$ (cm$^{-1}$) | $J_i$ | (Å) | (cm$^{-1}$) | BF[c] | unc. (%) | BF | unc. (%) |
| | | | | | Even-parity Upper Levels | | | | |
| 33398.70 | 3 | 0.00 | 4 | 2993.26 | 33398.70 | 0.234 | 1 | 0.237 | 3 |
| | | 236.95 | 3 | 3014.64 | 33161.75 | 0.673 | 1 | 0.668 | 0.3 |
| | | 8769.68 | 2 | 4059.11 | 24629.02 | 0.015 | 9 | 0.015 | 18 |
| | | 8957.47 | 3 | 4090.29 | 24441.23 | 0.034 | 9 | 0.031 | 13 |
| | | | | | residual | 0.045 | | 0.050 | |
| 35753.72 | 3 | 0.00 | 4 | 2796.09 | 35753.72 | 0.054 | 8 | ... | ... |
| | | 236.95 | 3 | 2814.74[d] | 35516.77 | 0.025 | 9 | ... | ... |
| | | 8769.68 | 2 | 3704.84 | 26984.04 | 0.707 | 1 | ... | ... |
| | | 8957.47 | 3 | 3730.81 | 26796.25 | 0.214 | 2 | ... | ... |
| | | | | | residual | 0.000 | | | |
| 35833.62 | 2 | 236.95 | 3 | 2808.42 | 35596.67 | 0.647 | 3 | ... | ... |
| | | 8769.68 | 2 | 3693.90 | 27063.94 | 0.046 | 12 | ... | ... |
| | | 8957.47 | 3 | 3719.71 | 26876.15 | 0.278 | 7 | ... | ... |
| | | | | | residual | 0.029 | | | |
| 36041.02 | 3 | 0.00 | 4 | 2773.80 | 36041.02 | 0.051 | 9 | ... | ... |
| | | 236.95 | 3 | 2792.16[d] | 35804.07 | 0.223 | 7 | ... | ... |
| | | 8769.68 | 2 | 3665.81 | 27271.34 | 0.697 | 2 | ... | ... |
| | | 8957.47 | 3 | 3691.23 | 27083.55 | 0.0240 | 4 | ... | ... |
| | | | | | residual | 0.005 | | | |
| 36132.08 | 3 | 0.00 | 4 | 2766.81 | 36132.08 | 0.014 | 9 | ... | ... |
| | | 236.95 | 3 | 2785.07 | 35895.13 | 0.174 | 7 | ... | ... |
| | | 8769.68 | 2 | 3653.61 | 27362.40 | 0.472 | 2 | ... | ... |
| | | 8957.47 | 3 | 3678.86 | 27174.61 | 0.340 | 2 | ... | ... |
| | | | | | residual | 0.000 | | | |
| 36394.64 | 2 | 236.95 | 3 | 2764.85 | 36157.69 | 0.0043 | 10 | ... | ... |
| | | 8769.68 | 2 | 3618.88 | 27624.96 | 0.0007 | 18 | ... | ... |
| | | 8957.47 | 3 | 3643.65 | 27437.17 | 0.995 | 0.3 | ... | ... |
| | | | | | residual | 0.000 | | | |
| 36547.84 | 4 | 0.00 | 4 | 2735.33 | 36547.84 | 0.195 | 3 | ... | ... |
| | | 236.95 | 3 | 2753.18 | 36310.89 | 0.575 | 2 | ... | ... |
| | | 8957.47 | 3 | 3623.42 | 27590.37 | 0.230 | 7 | ... | ... |
| | | | | | residual | 0.000 | | | |

| Upper Level[a] | | Lower Level[a] | | $\lambda_{air}$ | $\sigma_{vac}$ | This Study | | UW97[b] | |
|---|---|---|---|---|---|---|---|---|---|
| $E_k$ (cm$^{-1}$) | $J_k$ | $E_i$ (cm$^{-1}$) | $J_i$ | (Å) | (cm$^{-1}$) | BF[c] | unc. (%) | BF | unc. (%) |
| 37482.67 | 2 | 236.95 | 3 | 2684.08[d] | 37245.72 | 0.162 | 7 | ... | ... |
| | | 8769.68 | 2 | 3481.75 | 28712.99 | 0.742 | 1 | ... | ... |
| | | 8957.47 | 3 | 3504.67 | 28525.20 | 0.095 | 3 | ... | ... |
| | | | | residual | | 0.000 | | | |
| 37581.47 | 4 | 0.00 | 4 | 2660.09 | 37581.47 | 0.454 | 5 | ... | ... |
| | | 8957.47 | 3 | 3492.57 | 28624.00 | 0.541 | 4 | ... | ... |
| | | | | residual | | 0.005 | | | |
| 38093.53 | 4 | 0.00 | 4 | 2624.34 | 38093.53 | 0.397 | 5 | ... | ... |
| | | 236.95 | 3 | 2640.76 | 37856.58 | 0.099 | 5 | ... | ... |
| | | 8957.47 | 3 | 3431.19 | 29136.06 | 0.490 | 5 | ... | ... |
| | | 20228.75 | 5 | 5596.05 | 17864.78 | 0.008 | 29 | ... | ... |
| | | | | residual | | 0.006 | | | |
| 38361.24 | 4 | 0.00 | 4 | 2606.02 | 38361.24 | 0.162 | 8 | ... | ... |
| | | 236.95 | 3 | 2622.22 | 38124.29 | 0.054 | 9 | ... | ... |
| | | 8957.47 | 3 | 3399.95 | 29403.77 | 0.782 | 2 | ... | ... |
| | | | | residual | | 0.002 | | | |
| 38582.95 | 3 | 0.00 | 4 | 2591.04 | 38582.95 | 0.0033 | 13 | ... | ... |
| | | 236.95 | 3 | 2607.06 | 38346.00 | 0.395 | 5 | ... | ... |
| | | 8769.68 | 2 | 3353.25 | 29813.27 | 0.0217 | 4 | ... | ... |
| | | 8957.47 | 3 | 3374.50[d] | 29625.48 | 0.574 | 4 | ... | ... |
| | | | | residual | | 0.006 | | | |
| 39000.76 | 2 | 236.95 | 3 | 2578.95 | 38763.81 | 0.0037 | 23 | ... | ... |
| | | 8769.68 | 2 | 3306.90 | 30231.08 | 0.632 | 1 | ... | ... |
| | | 8957.47 | 3 | 3327.57 | 30043.29 | 0.365 | 1 | ... | ... |
| | | | | residual | | 0.000 | | | |
| 39162.07 | 4 | 0.00 | 4 | 2552.73 | 39162.07 | 0.021 | 11 | ... | ... |
| | | 236.95 | 3 | 2568.27 | 38925.12 | 0.0082 | 10 | ... | ... |
| | | 8957.47 | 3 | 3309.80 | 30204.60 | 0.971 | 0.3 | ... | ... |
| | | | | residual | | 0.000 | | | |
| 39196.70 | 2 | 236.95 | 3 | 2565.98 | 38959.75 | 0.019 | 9 | ... | ... |
| | | 8769.68 | 2 | 3285.61 | 30427.02 | 0.906 | 0.3 | ... | ... |
| | | 8957.47 | 3 | 3306.01[d] | 30239.23 | 0.074 | 5 | ... | ... |
| | | | | residual | | 0.001 | | | |

| Upper Level[a] | | Lower Level[a] | | $\lambda_{air}$ | $\sigma_{vac}$ | This Study | | UW97[b] | |
|---|---|---|---|---|---|---|---|---|---|
| $E_k$ (cm$^{-1}$) | $J_k$ | $E_i$ (cm$^{-1}$) | $J_i$ | (Å) | (cm$^{-1}$) | BF[c] | unc. (%) | BF | unc. (%) |
| 39514.68 | 2 | 236.95 | 3 | 2545.21 | 39277.73 | 0.018 | 14 | ... | ... |
| | | 8769.68 | 2 | 3251.62 | 30745.00 | 0.841 | 2 | ... | ... |
| | | 8957.47 | 3 | 3271.61 | 30557.21 | 0.043 | 27 | ... | ... |
| | | 17624.65 | 2 | 4567.01 | 21890.03 | 0.023 | 17 | ... | ... |
| | | 21978.77 | 2 | 5701.00 | 17535.91 | 0.075 | 18 | ... | ... |
| | | | | *residual* | | 0.000 | | | |
| 39554.05 | 3 | 0.00 | 4 | 2527.43 | 39554.05 | 0.019 | 9 | ... | ... |
| | | 236.95 | 3 | 2542.66 | 39317.10 | 0.024 | 9 | ... | ... |
| | | 8769.68 | 2 | 3247.47[d] | 30784.37 | 0.331 | 2 | ... | ... |
| | | 8957.47 | 3 | 3267.40 | 30596.58 | 0.616 | 1 | ... | ... |
| | | 21713.74 | 3 | 5603.73 | 17840.31 | 0.008 | 27 | ... | ... |
| | | | | *residual* | | 0.002 | | | |
| 39636.53 | 4 | 0.00 | 4 | 2522.17 | 39636.53 | 0.590 | 4 | ... | ... |
| | | 8957.47 | 3 | 3258.61 | 30679.06 | 0.390 | 6 | ... | ... |
| | | | | *residual* | | 0.020 | | | |
| 39843.24 | 4 | 0.00 | 4 | 2509.08 | 39843.24 | 0.303 | 6 | ... | ... |
| | | 236.95 | 3 | 2524.09 | 39606.29 | 0.0127 | 7 | ... | ... |
| | | 8957.47 | 3 | 3236.80 | 30885.77 | 0.676 | 3 | ... | ... |
| | | | | *residual* | | 0.008 | | | |
| 39893.87 | 3 | 0.00 | 4 | 2505.90 | 39893.87 | 0.016 | 10 | ... | ... |
| | | 236.95 | 3 | 2520.87 | 39656.92 | 0.054 | 8 | ... | ... |
| | | 8769.68 | 2 | 3212.01 | 31124.19 | 0.502 | 1 | ... | ... |
| | | 8957.47 | 3 | 3231.51 | 30936.40 | 0.428 | 1 | ... | ... |
| | | | | *residual* | | 0.000 | | | |
| 40056.32 | 2 | 236.95 | 3 | 2510.59 | 39819.37 | 0.007 | 15 | ... | ... |
| | | 8769.68 | 2 | 3195.33 | 31286.64 | 0.723 | 1 | ... | ... |
| | | 8957.47 | 3 | 3214.62[d] | 31098.85 | 0.218 | 1 | ... | ... |
| | | 22141.96 | 1 | 5580.57 | 17914.36 | 0.047 | 19 | ... | ... |
| | | | | *residual* | | 0.006 | | | |
| 40232.29 | 2 | 236.95 | 3 | 2499.54 | 39995.34 | 0.088 | 8 | ... | ... |
| | | 8769.68 | 2 | 3177.46 | 31462.61 | 0.495 | 1 | ... | ... |
| | | 8957.47 | 3 | 3196.54 | 31274.82 | 0.404 | 1 | ... | ... |
| | | 22141.96 | 1 | 5526.28 | 18090.33 | 0.014 | 26 | ... | ... |
| | | | | *residual* | | 0.000 | | | |

| Upper Level[a] | | Lower Level[a] | | $\lambda_{air}$ | $\sigma_{vac}$ | This Study | | UW97[b] | |
|---|---|---|---|---|---|---|---|---|---|
| $E_k$ (cm$^{-1}$) | $J_k$ | $E_i$ (cm$^{-1}$) | $J_i$ | (Å) | (cm$^{-1}$) | BF[c] | unc. (%) | BF | unc. (%) |
| 40359.75 | 3 | 0.00 | 4 | 2476.97 | 40359.75 | 0.206 | 2 | ... | ... |
| | | 236.95 | 3 | 2491.60 | 40122.80 | 0.627 | 2 | ... | ... |
| | | 8769.68 | 2 | 3164.64 | 31590.07 | 0.156 | 7 | ... | ... |
| | | | | | residual | 0.011 | | | |
| 40545.25 | 3 | 236.95 | 3 | 2480.13 | 40308.30 | 0.708 | 2 | ... | ... |
| | | 8769.68 | 2 | 3146.16 | 31775.57 | 0.190 | 7 | ... | ... |
| | | 8957.47 | 3 | 3164.87 | 31587.78 | 0.077 | 9 | ... | ... |
| | | 22457.51 | 4 | 5527.07 | 18087.74 | 0.0030 | 30 | ... | ... |
| | | 21978.77 | 2 | 5384.55 | 18566.48 | 0.007 | 21 | ... | ... |
| | | 23524.09 | 4 | 5873.41 | 17021.16 | 0.009 | 25 | ... | ... |
| | | 23934.73 | 3 | 6018.61 | 16610.52 | 0.006 | 23 | ... | ... |
| | | | | | residual | 0.000 | | | |

[a] Level energy and J values are taken from NIST ASD, with additions and corrections from Wyart (2011). The levels are ordered by parity and increasing energy. Ritz wavelengths are calculated from the energy levels using the index of air from Peck & Reeder (1972).

[b] UW97: Wickliffe & Lawler (1997), BFs are calculated from their *A*-values. Uncertainties are those quoted for their *A*-values reduced in quadrature by the lifetime uncertainty.

[c] BFs are given to three figures past the decimal except for cases where the absolute uncertainty is less than 0.001, in which case it is given to four places past the decimal. Occasionally in this table the BFs and residuals do not quite add to 1. This is due to rounding errors, but this discrepancy is well within the uncertainties.

[d] This transition was blended with another line in our spectra. The blend is with a line of Tm II, Tm I, Fe I or Fe II. A least-squares analysis was used to determine the blend fraction in each spectrum. See text for further discussion.

Table 4.
*A*-values and log(*gf*)s for 224 transitions of Tm II

| $\lambda_{air}$ (Å) | $E_k$ (cm$^{-1}$) | $J_k$ | $E_i$ (cm$^{-1}$) | $J_i$ | $A_{ki}$ (10$^6$ s$^{-1}$) | $\Delta A_{ki}$ (10$^6$ s$^{-1}$) | log(*gf*) |
|---|---|---|---|---|---|---|---|
| 2350.47 | 54988.96 | 7 | 12457.29 | 6 | 2.0 | 0.5 | -1.61 |
| 2421.65 | 53738.88 | 7 | 12457.29 | 6 | 21 | 3 | -0.56 |
| 2445.47 | 53336.89 | 5 | 12457.29 | 6 | 35.5 | 2.5 | -0.46 |
| 2447.38 | 53304.88 | 7 | 12457.29 | 6 | 11.7 | 1.0 | -0.80 |
| 2464.95 | 53013.80 | 6 | 12457.29 | 6 | 15.4 | 1.3 | -0.74 |
| 2476.97 | 40359.75 | 3 | 0.00 | 4 | 1.62 | 0.09 | -1.98 |
| 2480.13 | 40545.25 | 3 | 236.95 | 3 | 49 | 3 | -0.50 |
| 2491.60 | 40359.75 | 3 | 236.95 | 3 | 4.9 | 0.3 | -1.49 |
| 2499.54 | 40232.29 | 2 | 236.95 | 3 | 1.80 | 0.17 | -2.07 |
| 2505.90 | 39893.87 | 3 | 0.00 | 4 | 0.54 | 0.06 | -2.45 |
| 2507.14 | 52331.33 | 7 | 12457.29 | 6 | 24.7 | 2.1 | -0.46 |
| 2509.08 | 39843.24 | 4 | 0.00 | 4 | 37 | 3 | -0.51 |
| 2510.59 | 40056.32 | 2 | 236.95 | 3 | 0.18 | 0.03 | -3.08 |
| 2520.87 | 39893.87 | 3 | 236.95 | 3 | 1.87 | 0.18 | -1.90 |
| 2522.17 | 39636.53 | 4 | 0.00 | 4 | 4.2 | 0.3 | -1.44 |
| 2524.09 | 39843.24 | 4 | 236.95 | 3 | 1.53 | 0.13 | -1.88 |
| 2527.43 | 39554.05 | 3 | 0.00 | 4 | 1.51 | 0.16 | -2.00 |
| 2542.66 | 39554.05 | 3 | 236.95 | 3 | 1.91 | 0.20 | -1.89 |
| 2545.21 | 39514.68 | 2 | 236.95 | 3 | 0.30 | 0.04 | -2.84 |
| 2552.73 | 39162.07 | 4 | 0.00 | 4 | 1.20 | 0.14 | -1.98 |
| 2565.98 | 39196.70 | 2 | 236.95 | 3 | 1.37 | 0.14 | -2.17 |
| 2568.27 | 39162.07 | 4 | 236.95 | 3 | 0.47 | 0.05 | -2.38 |
| 2578.95 | 39000.76 | 2 | 236.95 | 3 | 0.059 | 0.014 | -3.53 |
| 2591.04 | 38582.95 | 3 | 0.00 | 4 | 0.19 | 0.03 | -2.87 |
| 2606.02 | 38361.24 | 4 | 0.00 | 4 | 3.4 | 0.3 | -1.50 |
| 2607.06 | 38582.95 | 3 | 236.95 | 3 | 23.0 | 1.6 | -0.79 |
| 2622.22 | 38361.24 | 4 | 236.95 | 3 | 1.15 | 0.12 | -1.97 |
| 2624.34 | 38093.53 | 4 | 0.00 | 4 | 19.4 | 1.4 | -0.74 |
| 2640.76 | 38093.53 | 4 | 236.95 | 3 | 4.8 | 0.3 | -1.34 |
| 2660.09 | 37581.47 | 4 | 0.00 | 4 | 4.5 | 0.3 | -1.36 |
| 2679.57 | 49765.68 | 7 | 12457.29 | 6 | 69 | 4 | 0.05 |
| 2684.08 | 37482.67 | 2 | 236.95 | 3 | 2.34 | 0.20 | -1.90 |
| 2718.85 | 53336.89 | 5 | 16567.47 | 5 | 2.8 | 0.5 | -1.46 |
| 2735.33 | 36547.84 | 4 | 0.00 | 4 | 0.47 | 0.03 | -2.33 |
| 2742.95 | 53013.80 | 6 | 16567.47 | 5 | 31.3 | 2.1 | -0.34 |
| 2753.18 | 36547.84 | 4 | 236.95 | 3 | 1.37 | 0.07 | -1.85 |

| $\lambda_{air}$ (Å) | $E_k$ (cm$^{-1}$) | $J_k$ | $E_i$ (cm$^{-1}$) | $J_i$ | $A_{ki}$ (10$^6$ s$^{-1}$) | $\Delta A_{ki}$ (10$^6$ s$^{-1}$) | log($gf$) |
|---|---|---|---|---|---|---|---|
| 2764.85 | 36394.64 | 2 | 236.95 | 3 | 0.194 | 0.022 | -2.95 |
| 2766.81 | 36132.08 | 3 | 0.00 | 4 | 0.61 | 0.06 | -2.31 |
| 2773.80 | 36041.02 | 3 | 0.00 | 4 | 0.86 | 0.09 | -2.16 |
| 2785.07 | 36132.08 | 3 | 236.95 | 3 | 7.5 | 0.6 | -1.21 |
| 2792.16 | 36041.02 | 3 | 236.95 | 3 | 3.7 | 0.3 | -1.52 |
| 2796.09 | 35753.72 | 3 | 0.00 | 4 | 0.85 | 0.08 | -2.16 |
| 2797.27 | 48195.94 | 6 | 12457.29 | 6 | 134 | 7 | 0.31 |
| 2808.42 | 35833.62 | 2 | 236.95 | 3 | 4.00 | 0.24 | -1.63 |
| 2814.74 | 35753.72 | 3 | 236.95 | 3 | 0.40 | 0.04 | -2.48 |
| 2818.47 | 47927.14 | 6 | 12457.29 | 6 | 43.3 | 2.4 | -0.17 |
| 2819.08 | 54988.96 | 7 | 19526.82 | 6 | 32 | 4 | -0.25 |
| 2826.43 | 54988.96 | 7 | 19619.08 | 7 | 33 | 3 | -0.22 |
| 2827.02 | 53336.89 | 5 | 17974.30 | 4 | 118 | 6 | 0.19 |
| 2827.92 | 47808.54 | 5 | 12457.29 | 6 | 137 | 7 | 0.26 |
| 2831.55 | 54988.96 | 7 | 19682.97 | 8 | 71 | 5 | 0.11 |
| 2852.59 | 53336.89 | 5 | 18291.37 | 4 | 9.0 | 0.6 | -0.92 |
| 2922.09 | 53738.88 | 7 | 19526.82 | 6 | 26.4 | 1.9 | -0.30 |
| 2929.99 | 53738.88 | 7 | 19619.08 | 7 | 4.5 | 0.5 | -1.06 |
| 2935.49 | 53738.88 | 7 | 19682.97 | 8 | 10.4 | 1.7 | -0.70 |
| 2935.99 | 50617.56 | 6 | 16567.47 | 5 | 207 | 11 | 0.54 |
| 2959.64 | 53304.88 | 7 | 19526.82 | 6 | 58 | 5 | 0.06 |
| 2967.74 | 53304.88 | 7 | 19619.08 | 7 | 38.1 | 2.2 | -0.12 |
| 2973.38 | 53304.88 | 7 | 19682.97 | 8 | 62 | 7 | 0.09 |
| 2985.36 | 53013.80 | 6 | 19526.82 | 6 | 47 | 3 | -0.09 |
| 2993.26 | 33398.70 | 3 | 0.00 | 4 | 2.58 | 0.13 | -1.62 |
| 2993.61 | 53013.80 | 6 | 19619.08 | 7 | 2.01 | 0.19 | -1.45 |
| 3014.64 | 33398.70 | 3 | 236.95 | 3 | 7.4 | 0.4 | -1.15 |
| 3047.48 | 52331.33 | 7 | 19526.82 | 6 | 13.4 | 1.0 | -0.55 |
| 3055.55 | 53336.89 | 5 | 20619.05 | 5 | 5.2 | 0.4 | -1.10 |
| 3056.07 | 52331.33 | 7 | 19619.08 | 7 | 130 | 7 | 0.44 |
| 3062.05 | 52331.33 | 7 | 19682.97 | 8 | 23.1 | 1.3 | -0.31 |
| 3063.44 | 54988.96 | 7 | 22355.43 | 6 | 4.8 | 1.0 | -0.99 |
| 3093.65 | 53336.89 | 5 | 21021.98 | 5 | 18.0 | 2.0 | -0.55 |
| 3124.89 | 53013.80 | 6 | 21021.98 | 5 | 35.8 | 2.1 | -0.17 |
| 3127.55 | 54988.96 | 7 | 23024.29 | 8 | 19.6 | 1.3 | -0.36 |
| 3146.16 | 40545.25 | 3 | 8769.68 | 2 | 13.2 | 1.1 | -0.86 |
| 3155.01 | 53738.88 | 7 | 22052.39 | 7 | 3.0 | 0.8 | -1.18 |

| $\lambda_{air}$ (Å) | $E_k$ (cm$^{-1}$) | $J_k$ | $E_i$ (cm$^{-1}$) | $J_i$ | $A_{ki}$ (10$^6$ s$^{-1}$) | $\Delta A_{ki}$ (10$^6$ s$^{-1}$) | log($gf$) |
|---|---|---|---|---|---|---|---|
| 3160.79 | 48195.94 | 6 | 16567.47 | 5 | 1.61 | 0.17 | -1.50 |
| 3164.64 | 40359.75 | 3 | 8769.68 | 2 | 1.23 | 0.11 | -1.89 |
| 3164.87 | 40545.25 | 3 | 8957.47 | 3 | 5.3 | 0.5 | -1.25 |
| 3173.97 | 51116.29 | 8 | 19619.08 | 7 | 4.2 | 0.3 | -0.96 |
| 3177.46 | 40232.29 | 2 | 8769.68 | 2 | 10.1 | 0.5 | -1.12 |
| 3180.42 | 51116.29 | 8 | 19682.97 | 8 | 12.6 | 0.7 | -0.49 |
| 3185.47 | 53738.88 | 7 | 22355.43 | 6 | 21 | 6 | -0.31 |
| 3195.33 | 40056.32 | 2 | 8769.68 | 2 | 17.6 | 0.9 | -0.87 |
| 3196.54 | 40232.29 | 2 | 8957.47 | 3 | 8.2 | 0.4 | -1.20 |
| 3198.82 | 53304.88 | 7 | 22052.39 | 7 | 9.6 | 0.7 | -0.66 |
| 3199.99 | 47808.54 | 5 | 16567.47 | 5 | 18.7 | 1.3 | -0.50 |
| 3212.01 | 39893.87 | 3 | 8769.68 | 2 | 17.3 | 0.9 | -0.73 |
| 3214.62 | 40056.32 | 2 | 8957.47 | 3 | 5.3 | 0.3 | -1.39 |
| 3215.46 | 50617.56 | 6 | 19526.82 | 6 | 5.2 | 0.5 | -0.98 |
| 3216.11 | 54988.96 | 7 | 23904.43 | 6 | 40 | 3 | -0.03 |
| 3221.96 | 53336.89 | 5 | 22308.82 | 4 | 3.3 | 0.4 | -1.24 |
| 3222.03 | 54988.96 | 7 | 23961.60 | 7 | 24.7 | 1.6 | -0.24 |
| 3226.81 | 53336.89 | 5 | 22355.43 | 6 | 37.1 | 2.4 | -0.20 |
| 3228.90 | 53013.80 | 6 | 22052.39 | 7 | 35.7 | 2.1 | -0.14 |
| 3230.14 | 53304.88 | 7 | 22355.43 | 6 | 21.1 | 1.4 | -0.30 |
| 3231.51 | 39893.87 | 3 | 8957.47 | 3 | 14.8 | 0.8 | -0.79 |
| 3232.19 | 54988.96 | 7 | 24059.08 | 8 | 2.6 | 0.3 | -1.22 |
| 3236.80 | 39843.24 | 4 | 8957.47 | 3 | 81 | 5 | 0.06 |
| 3247.47 | 39554.05 | 3 | 8769.68 | 2 | 26.0 | 1.4 | -0.54 |
| 3251.62 | 39514.68 | 2 | 8769.68 | 2 | 14.2 | 0.7 | -0.95 |
| 3254.84 | 53738.88 | 7 | 23024.29 | 8 | 18.5 | 1.2 | -0.36 |
| 3258.61 | 39636.53 | 4 | 8957.47 | 3 | 2.77 | 0.22 | -1.40 |
| 3260.81 | 53013.80 | 6 | 22355.43 | 6 | 3.1 | 0.3 | -1.20 |
| 3261.65 | 51116.29 | 8 | 20465.82 | 9 | 117 | 6 | 0.50 |
| 3267.40 | 39554.05 | 3 | 8957.47 | 3 | 48.5 | 2.5 | -0.26 |
| 3271.61 | 39514.68 | 2 | 8957.47 | 3 | 0.73 | 0.20 | -2.23 |
| 3285.61 | 39196.70 | 2 | 8769.68 | 2 | 65 | 3 | -0.28 |
| 3301.50 | 53304.88 | 7 | 23024.29 | 8 | 20.3 | 1.5 | -0.30 |
| 3301.68 | 52331.33 | 7 | 22052.39 | 7 | 10.1 | 0.7 | -0.60 |
| 3306.01 | 39196.70 | 2 | 8957.47 | 3 | 5.3 | 0.4 | -1.36 |
| 3306.90 | 39000.76 | 2 | 8769.68 | 2 | 10.1 | 0.5 | -1.08 |
| 3309.80 | 39162.07 | 4 | 8957.47 | 3 | 55 | 3 | -0.09 |

| $\lambda_{air}$ (Å) | $E_k$ (cm$^{-1}$) | $J_k$ | $E_i$ (cm$^{-1}$) | $J_i$ | $A_{ki}$ (10$^6$ s$^{-1}$) | $\Delta A_{ki}$ (10$^6$ s$^{-1}$) | log($gf$) |
|---|---|---|---|---|---|---|---|
| 3316.17 | 49765.68 | 7 | 19619.08 | 7 | 17.6 | 1.7 | -0.36 |
| 3323.21 | 49765.68 | 7 | 19682.97 | 8 | 29.5 | 2.5 | -0.14 |
| 3327.57 | 39000.76 | 2 | 8957.47 | 3 | 5.8 | 0.3 | -1.31 |
| 3335.05 | 52331.33 | 7 | 22355.43 | 6 | 17.8 | 1.3 | -0.35 |
| 3350.87 | 53738.88 | 7 | 23904.43 | 6 | 8.4 | 1.1 | -0.68 |
| 3353.25 | 38582.95 | 3 | 8769.68 | 2 | 1.26 | 0.08 | -1.83 |
| 3357.30 | 53738.88 | 7 | 23961.60 | 7 | 12.0 | 1.4 | -0.52 |
| 3368.33 | 53738.88 | 7 | 24059.08 | 8 | 54 | 3 | 0.14 |
| 3374.50 | 38582.95 | 3 | 8957.47 | 3 | 33.4 | 2.1 | -0.40 |
| 3377.91 | 50617.56 | 6 | 21021.98 | 5 | 13.9 | 1.0 | -0.51 |
| 3381.06 | 53336.89 | 5 | 23768.84 | 5 | 8.9 | 1.0 | -0.78 |
| 3396.63 | 53336.89 | 5 | 23904.43 | 6 | 11.1 | 1.0 | -0.68 |
| 3399.95 | 38361.24 | 4 | 8957.47 | 3 | 16.5 | 0.9 | -0.59 |
| 3400.33 | 53304.88 | 7 | 23904.43 | 6 | 8.6 | 0.6 | -0.65 |
| 3418.41 | 53013.80 | 6 | 23768.84 | 5 | 17.9 | 1.3 | -0.39 |
| 3431.19 | 38093.53 | 4 | 8957.47 | 3 | 23.9 | 1.7 | -0.42 |
| 3439.71 | 51116.29 | 8 | 22052.39 | 7 | 4.6 | 0.3 | -0.86 |
| 3441.09 | 53013.80 | 6 | 23961.60 | 7 | 15.4 | 1.3 | -0.45 |
| 3481.75 | 37482.67 | 2 | 8769.68 | 2 | 10.7 | 0.5 | -1.01 |
| 3487.08 | 48195.94 | 6 | 19526.82 | 6 | 26.6 | 2.5 | -0.20 |
| 3492.57 | 37581.47 | 4 | 8957.47 | 3 | 5.4 | 0.3 | -1.05 |
| 3499.77 | 50617.56 | 6 | 22052.39 | 7 | 7.5 | 0.7 | -0.75 |
| 3504.67 | 37482.67 | 2 | 8957.47 | 3 | 1.37 | 0.08 | -1.90 |
| 3520.08 | 47927.14 | 6 | 19526.82 | 6 | 5.4 | 0.7 | -0.89 |
| 3523.88 | 52331.33 | 7 | 23961.60 | 7 | 1.2 | 0.3 | -1.47 |
| 3529.80 | 53336.89 | 5 | 25014.75 | 5 | 7.9 | 0.8 | -0.79 |
| 3534.84 | 47808.54 | 5 | 19526.82 | 6 | 61 | 4 | 0.10 |
| 3536.03 | 52331.33 | 7 | 24059.08 | 8 | 13.0 | 1.0 | -0.44 |
| 3537.29 | 50617.56 | 6 | 22355.43 | 6 | 3.4 | 0.4 | -1.08 |
| 3558.72 | 51116.29 | 8 | 23024.29 | 8 | 16.4 | 1.1 | -0.28 |
| 3607.35 | 49765.68 | 7 | 22052.39 | 7 | 31 | 3 | -0.05 |
| 3618.88 | 36394.64 | 2 | 8769.68 | 2 | 0.030 | 0.006 | -3.52 |
| 3623.42 | 36547.84 | 4 | 8957.47 | 3 | 0.55 | 0.05 | -2.01 |
| 3637.63 | 53738.88 | 7 | 26256.26 | 7 | 15.9 | 1.1 | -0.33 |
| 3643.65 | 36394.64 | 2 | 8957.47 | 3 | 45.2 | 2.3 | -0.35 |
| 3647.23 | 49765.68 | 7 | 22355.43 | 6 | 36 | 4 | 0.04 |
| 3653.61 | 36132.08 | 3 | 8769.68 | 2 | 20.4 | 1.1 | -0.54 |

| $\lambda_{air}$ (Å) | $E_k$ (cm$^{-1}$) | $J_k$ | $E_i$ (cm$^{-1}$) | $J_i$ | $A_{ki}$ (10$^6$ s$^{-1}$) | $\Delta A_{ki}$ (10$^6$ s$^{-1}$) | log($gf$) |
|---|---|---|---|---|---|---|---|
| 3660.88 | 47927.14 | 6 | 20619.05 | 5 | 12.8 | 1.1 | -0.48 |
| 3665.81 | 36041.02 | 3 | 8769.68 | 2 | 11.6 | 0.6 | -0.78 |
| 3667.25 | 53738.88 | 7 | 26478.28 | 6 | 23.6 | 2.3 | -0.15 |
| 3676.85 | 47808.54 | 5 | 20619.05 | 5 | 0.61 | 0.13 | -1.87 |
| 3677.98 | 39638.41 | 7 | 12457.29 | 6 | 12.5 | 0.9 | -0.42 |
| 3678.86 | 36132.08 | 3 | 8957.47 | 3 | 14.6 | 0.8 | -0.68 |
| 3678.95 | 48195.94 | 6 | 21021.98 | 5 | 35 | 4 | -0.04 |
| 3691.23 | 36041.02 | 3 | 8957.47 | 3 | 0.40 | 0.03 | -2.24 |
| 3693.90 | 35833.62 | 2 | 8769.68 | 2 | 0.28 | 0.04 | -2.54 |
| 3694.82 | 51116.29 | 8 | 24059.08 | 8 | 55 | 5 | 0.28 |
| 3704.84 | 35753.72 | 3 | 8769.68 | 2 | 11.1 | 0.6 | -0.80 |
| 3715.70 | 47927.14 | 6 | 21021.98 | 5 | 2.5 | 0.3 | -1.18 |
| 3719.71 | 35833.62 | 2 | 8957.47 | 3 | 1.72 | 0.15 | -1.75 |
| 3723.51 | 50617.56 | 6 | 23768.84 | 5 | 1.5 | 0.3 | -1.40 |
| 3726.58 | 53304.88 | 7 | 26478.28 | 6 | 6.4 | 0.6 | -0.70 |
| 3730.81 | 35753.72 | 3 | 8957.47 | 3 | 3.36 | 0.18 | -1.31 |
| 3732.15 | 47808.54 | 5 | 21021.98 | 5 | 6.9 | 0.7 | -0.80 |
| 3735.55 | 53336.89 | 5 | 26574.66 | 4 | 3.5 | 0.5 | -1.10 |
| 3736.20 | 53013.80 | 6 | 26256.26 | 7 | 10.0 | 2.2 | -0.57 |
| 3738.46 | 49765.68 | 7 | 23024.29 | 8 | 3.4 | 0.5 | -0.97 |
| 3754.45 | 53336.89 | 5 | 26709.43 | 6 | 2.4 | 0.4 | -1.25 |
| 3758.97 | 53304.88 | 7 | 26709.43 | 6 | 3.9 | 0.5 | -0.90 |
| 3767.46 | 53013.80 | 6 | 26478.28 | 6 | 6.4 | 0.6 | -0.75 |
| 3823.95 | 48195.94 | 6 | 22052.39 | 7 | 2.0 | 0.3 | -1.25 |
| 3833.99 | 52331.33 | 7 | 26256.26 | 7 | 6.0 | 1.5 | -0.70 |
| 3865.69 | 49765.68 | 7 | 23904.43 | 6 | 3.3 | 0.4 | -0.96 |
| 3866.92 | 52331.33 | 7 | 26478.28 | 6 | 5.9 | 0.6 | -0.70 |
| 3888.95 | 49765.68 | 7 | 24059.08 | 8 | 10.2 | 1.3 | -0.46 |
| 3901.81 | 52331.33 | 7 | 26709.43 | 6 | 1.9 | 0.5 | -1.18 |
| 3904.72 | 50617.56 | 6 | 25014.75 | 5 | 5.9 | 0.8 | -0.76 |
| 3960.73 | 53336.89 | 5 | 28096.17 | 6 | 1.6 | 0.4 | -1.38 |
| 4021.39 | 51116.29 | 8 | 26256.26 | 7 | 9.9 | 0.8 | -0.39 |
| 4059.11 | 33398.70 | 3 | 8769.68 | 2 | 0.162 | 0.017 | -2.55 |
| 4090.29 | 33398.70 | 3 | 8957.47 | 3 | 0.38 | 0.04 | -2.18 |
| 4103.71 | 50617.56 | 6 | 26256.26 | 7 | 9.8 | 1.2 | -0.49 |
| 4161.55 | 47927.14 | 6 | 23904.43 | 6 | 0.70 | 0.09 | -1.63 |
| 4182.20 | 47808.54 | 5 | 23904.43 | 6 | 0.76 | 0.15 | -1.66 |

| $\lambda_{air}$ (Å) | $E_k$ (cm$^{-1}$) | $J_k$ | $E_i$ (cm$^{-1}$) | $J_i$ | $A_{ki}$ (10$^6$ s$^{-1}$) | $\Delta A_{ki}$ (10$^6$ s$^{-1}$) | log($gf$) |
|---|---|---|---|---|---|---|---|
| 4224.42 | 54988.96 | 7 | 31323.73 | 6 | 0.72 | 0.14 | -1.54 |
| 4252.42 | 49765.68 | 7 | 26256.26 | 7 | 6.1 | 0.9 | -0.61 |
| 4292.96 | 49765.68 | 7 | 26478.28 | 6 | 5.7 | 0.9 | -0.62 |
| 4312.63 | 48195.94 | 6 | 25014.75 | 5 | 2.2 | 0.3 | -1.11 |
| 4336.00 | 49765.68 | 7 | 26709.43 | 6 | 3.3 | 0.5 | -0.85 |
| 4357.25 | 53304.88 | 7 | 30361.06 | 6 | 3.4 | 0.5 | -0.84 |
| 4556.67 | 48195.94 | 6 | 26256.26 | 7 | 8.0 | 1.3 | -0.49 |
| 4567.01 | 39514.68 | 2 | 17624.65 | 2 | 0.38 | 0.07 | -2.23 |
| 4613.20 | 47927.14 | 6 | 26256.26 | 7 | 3.8 | 0.8 | -0.80 |
| 4711.73 | 47927.14 | 6 | 26709.43 | 6 | 3.1 | 0.6 | -0.87 |
| 4970.87 | 39638.41 | 7 | 19526.82 | 6 | 2.38 | 0.17 | -0.88 |
| 4973.58 | 48195.94 | 6 | 28095.31 | 7 | 4.0 | 0.8 | -0.72 |
| 4993.78 | 39638.41 | 7 | 19619.08 | 7 | 1.53 | 0.14 | -1.07 |
| 5004.30 | 53013.80 | 6 | 33036.56 | 5 | 14 | 4 | -0.17 |
| 5009.77 | 39638.41 | 7 | 19682.97 | 8 | 10.9 | 0.6 | -0.21 |
| 5040.99 | 47927.14 | 6 | 28095.31 | 7 | 5.0 | 0.8 | -0.61 |
| 5041.00 | 53013.80 | 6 | 33182.00 | 7 | 7.3 | 1.3 | -0.44 |
| 5269.44 | 53738.88 | 7 | 34766.81 | 8 | 12 | 3 | -0.13 |
| 5384.55 | 40545.25 | 3 | 21978.77 | 2 | 0.49 | 0.10 | -1.83 |
| 5461.95 | 52331.33 | 7 | 34027.94 | 8 | 17 | 3 | 0.05 |
| 5526.28 | 40232.29 | 2 | 22141.96 | 1 | 0.28 | 0.07 | -2.19 |
| 5527.07 | 40545.25 | 3 | 22457.51 | 4 | 0.22 | 0.07 | -2.15 |
| 5574.36 | 51116.29 | 8 | 33182.00 | 7 | 5.2 | 1.0 | -0.39 |
| 5580.57 | 40056.32 | 2 | 22141.96 | 1 | 1.14 | 0.22 | -1.58 |
| 5589.93 | 54988.96 | 7 | 37104.63 | 8 | 16 | 4 | 0.06 |
| 5596.05 | 38093.53 | 4 | 20228.75 | 5 | 0.39 | 0.12 | -1.78 |
| 5603.73 | 39554.05 | 3 | 21713.74 | 3 | 0.64 | 0.18 | -1.67 |
| 5684.76 | 39638.41 | 7 | 22052.39 | 7 | 5.3 | 0.5 | -0.42 |
| 5691.21 | 47927.14 | 6 | 30361.06 | 6 | 4.9 | 1.0 | -0.51 |
| 5701.00 | 39514.68 | 2 | 21978.77 | 2 | 1.26 | 0.24 | -1.51 |
| 5733.82 | 50617.56 | 6 | 33182.00 | 7 | 12.3 | 2.5 | -0.10 |
| 5778.82 | 47808.54 | 5 | 30508.76 | 4 | 3.7 | 1.1 | -0.69 |
| 5784.44 | 39638.41 | 7 | 22355.43 | 6 | 0.71 | 0.11 | -1.27 |
| 5833.98 | 52331.33 | 7 | 35195.12 | 7 | 5.6 | 1.6 | -0.37 |
| 5850.32 | 51116.29 | 8 | 34027.94 | 8 | 3.3 | 1.0 | -0.54 |
| 5873.41 | 40545.25 | 3 | 23524.09 | 4 | 0.65 | 0.17 | -1.63 |
| 6018.61 | 40545.25 | 3 | 23934.73 | 3 | 0.40 | 0.10 | -1.81 |

| $\lambda_{air}$ (Å) | $E_k$ (cm$^{-1}$) | $J_k$ | $E_i$ (cm$^{-1}$) | $J_i$ | $A_{ki}$ (10$^6$ s$^{-1}$) | $\Delta A_{ki}$ (10$^6$ s$^{-1}$) | log($gf$) |
|---|---|---|---|---|---|---|---|
| 6076.98 | 54988.96 | 7 | 38537.96 | 6 | 6.6 | 2.0 | -0.26 |
| 6377.09 | 39638.41 | 7 | 23961.60 | 7 | 0.31 | 0.05 | -1.55 |
| 6416.99 | 39638.41 | 7 | 24059.08 | 8 | 0.37 | 0.11 | -1.46 |